\documentclass[12pt,onecolumn]{article}
\usepackage{color}
\usepackage{amsmath}
\usepackage{amssymb}
\usepackage{amsfonts}
\usepackage{geometry}
\usepackage{setspace}
\usepackage{amsmath,bm}
\usepackage{graphicx}
\usepackage[section]{placeins}
\usepackage{subfigure}
\usepackage[sort&compress]{natbib}

\author{Shuwei Zhou$^{1,2}$, Xiaoying Zhuang$^2$, Timon Rabczuk$^{3,4*}$}
\title {A phase-field modeling approach of fracture propagation in poroelastic media}
\begin{document}

\bibliographystyle{unsrtnat}
\setcitestyle{numbers,square,aysep={},yysep={,}}
\date{}
\maketitle

\spacing {1.2}
\noindent
1 Institute of Structural Mechanics, Bauhaus-University Weimar, Weimar 99423, Germany\\
2 Department of Geotechnical Engineering, College of Civil Engineering, Tongji University, Shanghai 200092, P.R. China\\
3 Division of Computational Mechanics, Ton Duc Thang University, Ho Chi Minh City, Viet Nam\\
4 Faculty of Civil Engineering, Ton Duc Thang University, Ho Chi Minh City, Viet Nam\\
* Corresponding author: Timon Rabczuk

\begin{abstract}
\noindent This paper proposes a phase field model for fracture in poroelastic media. The porous medium is modeled based on the classical Biot poroelasticity theory and the fracture behavior is controlled by the phase field model. Moreover, the fracture propagation is driven by the elastic energy where the phase field is used as an interpolation function to transit fluid property from the intact medium to the fully broken one. We use a segregated (staggered) scheme and implement our approach in Comsol Multiphysics. The proposed model is verified by a single-phase solid subjected to tension and a 2D specimen subjected to an increasing internal pressure. We also compare our results with analytical solutions. Finally, we show 2D and 3D examples of  internal fluid injection to illustrate the capability of the proposed approach.  
\end{abstract}

\noindent Keywords: Phase field, Hydraulic fractures, Poroelasticity, Comsol

\section {Introduction}
Fracture in poroelastic media is of major importance in mechanical, energy and environmental engineering \citep{mikelic2013phase, liu2014situ}.  In particular, predicting fracture propagation during water injection is critical for hydraulic fracturing (fracking), an interesting technique used to extract petroleum and natural gas (e.g. shale gas). Fractures are created by fracking to connect wellbores with expected petroleum or gas. Thus, vast amounts of resources, which are inaccessible in previous decades, will be extracted promising great economic benefits. However, fracking may also result in leakage of fracturing fluid or gas, which might contaminate the groundwater because of the unintentionally created channels \citep{vidic2013impact, ren2017equivalent}. Other reasons for objecting fracking are risks to air quality loss and surface contamination from spills \citep{mikelic2013phase}. Therefore, the vast economic benefits and continued controversies require accurate mathematical models and proper numerical simulation tools to treat complex problems on fracture propagation in poroelastic media \citep{figueiredo2017effects}. 

Fracture models for single-phasic solids can be classified into two categories, i.e. discrete and continuous approaches. Discrete approaches introduce discontinuities which capture the jump in the displacement field. Among the most popular computational methods for fracture are the extended finite element method (XFEM) \citep{moes2002extended}, cohesive elements, element-erosion techniques \citep{johnson1987eroding} and remeshing techniques \citep{areias2017steiner}. Discrete crack methods are usually computationally challenging and require complex crack tracking procedures. Furthermore, the lack of reliable cracking criteria reduces the robustness and reliability of discrete crack propagation, at least when they are based on a continuous crack topology. The cracking particle method \citep{rabczuk2004cracking}, peridynamics \citep{oterkus2017fully} and dual-horizon peridynamics \citep{ren2016dual} are discrete crack approaches which do not exhibit such drawbacks and are therefore well suited for dynamic fracture including complex crack interactions. They do not model the crack as continuous surface which also significantly facilitates the implementation. However, fluid flow through the opening of discrete cracks cannot be captured in such method due to the absence of crack path continuity. There are a few contributions modeling fracture through discrete cracks in extended finite element and meshfree methods, see e.g. the contributions in \citep{rabczuk2010immersed}. However, they are commonly restricted to a few number of cracks. Different from the discrete approaches are continuous approaches to fracture which do not introduce discontinuities in the displacement field. Popular approaches include gradient damage models \citep{peerlings1996some}, screened-poisson models \citep{areias2016damage} and phase field models \citep{miehe2010thermodynamically}. All these models introduce an intrinsic length scale into the discretization and smear the fracture over a localization band of finite width drastically facilitating their implementation.
 
To date, there are a few approaches of discrete and continuous approaches for hydraulic fracturing (HF). For example, \citet{wu2015simultaneous} modeled 2D crack propagation using the BEM. \citet{lecampion2009extended} applied the extended finite element method to hydraulic fracturing. The fluid pressure was applied along a line fracture in a 2D impermeable medium to drive the crack. However, the complexity of discrete approaches for HF remains high and studies are limited to fairly simple cases, mostly in 2D. PFMs seem to be a valuable alternative, especially since they can be easily extended to coupled problems. PFMs for HF have been proposed for instance in \citep{bourdin2012variational,wheeler2014augmented,miehe2015minimization,mikelic2015quasi, mikelic2015phase,heister2015primal, lee2016pressure,wick2016fluid, yoshioka2016variational,miehe2016phase,ehlers2017phase,santillan2017phase}. \citet{wheeler2014augmented} rewrote the energy functional by including poroelastic terms and succeeded in extending the phase field model to porous media. However, the variation in the reservoir and fracture domains with time is treated as a moving boundary problem and thereby extra work is needed for implementation. Later, \citet{mikelic2015quasi} fully coupled the three fields: elasticity, phase field, and pressure. They modified the energy functional from their previous work,  the flow in their entire system is governed by Biot equations. The permeability tensor was also modified to consider a higher permeability along the fracture. The implementation approaches of \citet{mikelic2015quasi} were then enhanced using adaptive element schemes by \citep{heister2015primal}. \citet{yoshioka2016variational} proposed approaches to couple the phase field model to reservoir simulators. Miehe et al. \citep{miehe2015minimization, miehe2016phase} proposed new minimization and saddle point principles for Darcy-Biot-type flow in fractured poroelastic media coupled with phase field modeling. The evolution of the phase field was driven by the effective stress in the solid skeleton and a stress threshold was set. Moreover, the flow in the fractures was set as Poiseuille-type by modeling Darcy flow with an anisotropic permeability tensor. Recently, \citet{ehlers2017phase} embedded a phase-field approach in the theory of porous media to model dynamic hydraulic fracturing. \citet{santillan2017phase} proposed an immersed-fracture formulation for impermeable porous media. However, these recently developed approaches for extending PFM to HF are quite complicated and computationally expensive.

In this paper another attempt for phase field modeling of fluid-driven fractures is made. The porous medium is modeled based on the classical Biot poroelasticity theory and the fracture behavior is controlled by the phase field model. We add an additional pressure-related term in the energy functional and derive the governing equations of the strong form. Different from Miehe et al. \citep{miehe2015minimization,miehe2016phase}, the fracture propagation is driven by the elastic energy and we set no stress threshold in the formulation. In addition, the phase field is used as an interpolation function to transit fluid property from the intact medium to the fully broken one. We use the FEM software Comsol Multiphysics to implement the proposed approach. A segregated (staggered) scheme is used and thereby the displacement, pressure, and phase field are solved independently. We verify our approach for two cases: a single-phasic solid and a 2D specimen subjected to an increasing internal pressure. In the second case, the proposed approach is compared with analytical solutions. Subsequently, more complex examples are studied.

The paper is organized as follows. We give the mathematical models for fracture in Section 2 followed by Section 3 which shows the numerical implementation of the phase field model in Comsol. In Section 4, we then verify the numerical simulation by two cases:  a single-phasic solid subjected to tension and a 2D specimen subjected to an increasing internal pressure. Afterwards, we give 2D and 3D examples of  internal fluid injection in Section 5. Finally, we end with conclusions regarding our work in Section 6. 

\section {Mathematical models for fracture in porous media}
\subsection {Theory of brittle fracture}

Let us consider a bounded domain $\Omega\subset \mathbb R^d$ ($d\in \{2,3\} $) with external boundary $\partial \Omega$ and an internal discontinuity boundary $\Gamma $. Furthermore, we assume: 

	\begin{spacing}{1.0}
	\begin{itemize}
	\item The intrinsic length scale parameter of the phase field is larger than the pore size.
	\item The porous medium is linear elastic, homogeneous, and isotropic.
	\item The fluid in the medium is compressible and viscous.
	\end{itemize}
	\end{spacing}

Let $T>0$ be the computational time interval and the displacement of body $\Omega $ at time $t\in[0,T]$ is denoted by $\bm u(\bm x,t)\subset \mathbb R^d$ where $\bm x $ is the position vector. The displacement field satisfies the time-dependent Dirichlet boundary conditions, $u_i(\bm x,t)=g_i(\bm x,t)$, on $\partial \Omega_{g_i} \in \Omega$, and the time-dependent Neumann conditions on $\partial \Omega_{h_i} \in \Omega$. The body $\Omega$ is subjected to a body force $\bm b(\bm x,t)\subset \mathbb R^d$ and a traction $\bm f(\bm x,t)$ acts on the the boundary $\partial \Omega_{h_i}$.

Griffith's theory \citep{francfort1998revisiting} assumes the energy to create a fracture surface per unit area is equal to the critical fracture energy density $G_c$, which is also commonly referred to as the critical energy release rate. Thus, for a pure single-phasic solid, the energy functional for the solid $\Psi(\bm u,\Gamma)$ can be expressed in terms of the elastic energy $\psi_{\varepsilon}(\bm \varepsilon)$, fracture energy and work from external forces. However, for a porous solid, the fluid pressure must be considered for the solid and an additional pressure-related term is used \citep{mikelic2015quasi, mikelic2015phase,lee2016pressure}. In this paper, we follow the idea of \citet{lee2016pressure} and rewrite the energy functional as
	\begin{equation}
	\Psi(\bm u,\Gamma) = \int_{\Omega}\psi_{\varepsilon}(\bm \varepsilon) d{\Omega}-\int_{\Omega}\alpha p \cdot (\nabla \cdot \bm u) d{\Omega}+\int_{\Gamma}G_c dS-\int_{\Omega} \bm b\cdot{\bm u}d{\Omega} - \int_{\partial\Omega_{h_i}} \bm f\cdot{\bm u}dS
	\label{functional2}
	\end{equation}

\noindent where $p:\Omega\times[0,T]\rightarrow\mathbb R$ is the fluid pressure, $\alpha\in [\epsilon_p,1]$ is the Biot coefficient and $\epsilon_p$ and
	\begin{equation}
	\varepsilon_{ij}=\frac 1 2 \left(\frac{\partial u_i}{\partial x_j}+\frac{\partial u_j}{\partial x_i}\right)
	\end{equation}
	
\noindent is the linear strain tensor. For an isotropic and linear elastic solid, the elastic energy density $\psi_{\varepsilon}(\bm \varepsilon)$ is given by \citep{miehe2010phase}
	\begin{equation}
	\psi_{\varepsilon}(\bm \varepsilon) = \frac{1}{2}\lambda\varepsilon_{ii}\varepsilon_{jj}+\mu\varepsilon_{ij}\varepsilon_{ij}
	\end{equation}

\noindent where $\lambda,\mu>0$ are the Lam\'e constants.

\subsection{Phase field approximation for fracture}

\begin{figure}[htbp]
	\centering
	\subfigure[]{\includegraphics[height = 5cm]{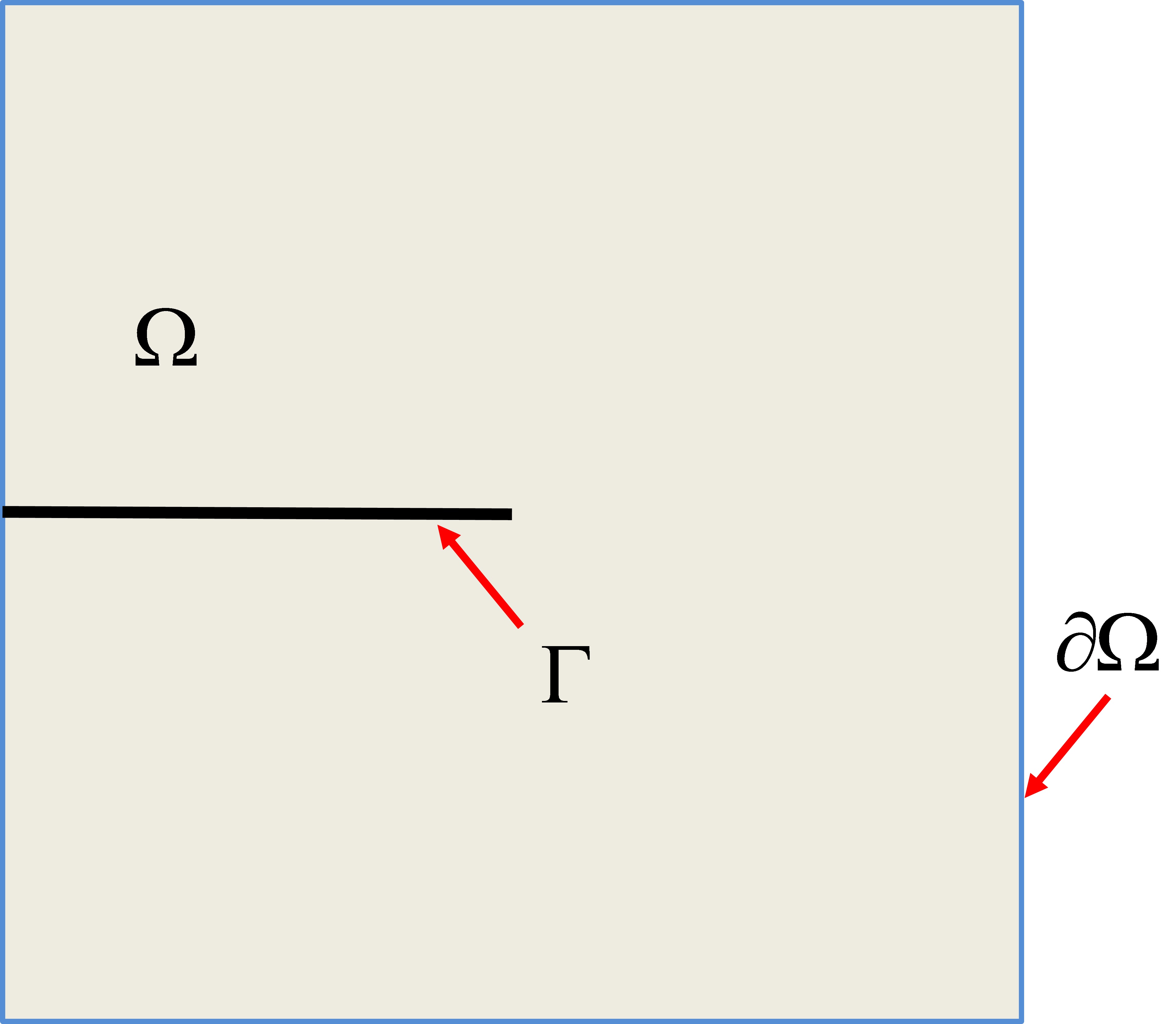}}
	\subfigure[]{\includegraphics[height = 5cm]{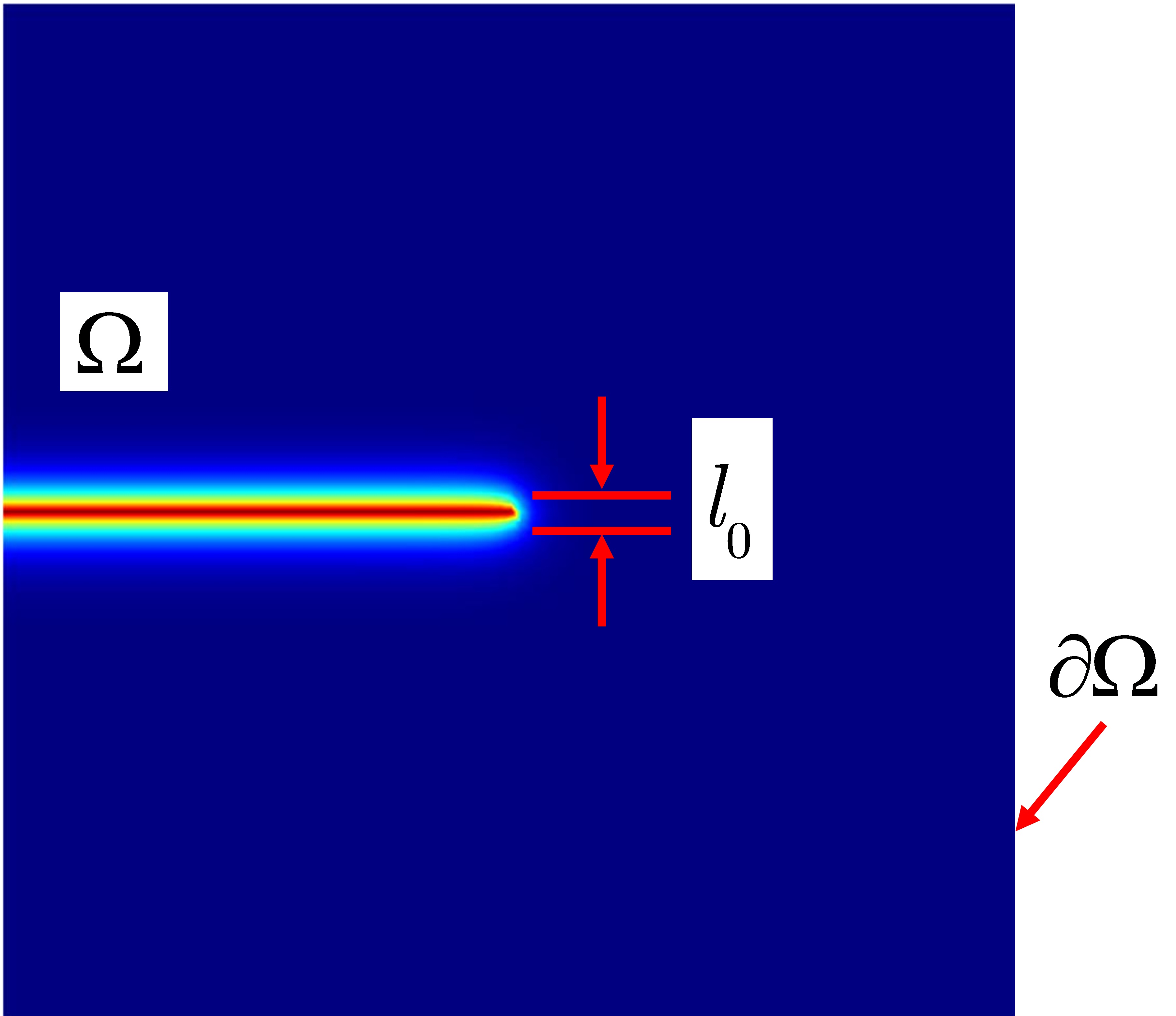}}
	\caption{Sharp and diffusive crack shape. (a) A sharp crack $\Gamma$ and (b) the diffusive phase field.}
	\label{Sharp and diffusive crack shape}
	\end{figure}

The  phase field $\phi(\bm x,t)\in[0,1]$ diffuses the fracture surface $\Gamma$ over a volume which avoids complex crack tracking procedures. $\phi=1$ indicates the 'fully' cracked region while $\phi=0$ denotes the unbroken material. A length scale parameter $l_0$ is used to control the transition region where the phase field varies from 0 to 1, thus,  reflecting the shape of a diffusive crack (see Fig. \ref{Sharp and diffusive crack shape} for details). The width of the crack region will increase with increasing $l_0$. It can be shown that the crack surface density per unit volume of the solid is  given by \cite{miehe2010phase}	
	\begin{equation}
	\gamma(\phi,\bigtriangledown\phi)=\frac{\phi^2}{2l_0}+\frac{l_0}2\frac{\partial\phi}{\partial x_i}\frac{\partial\phi}{\partial x_i}
	\label{phase field approximation}
	\end{equation}

By using Eq. \eqref{phase field approximation}, the fracture energy in Eq. \eqref{functional2} is calculated by
	\begin{equation}
	\int_{\Gamma}G_c dS=\int_{\Omega}G_c\left[\frac{\phi^2}{2l_0}+\frac{l_0}2\frac{\partial\phi}{\partial x_i}\frac{\partial\phi}{\partial x_i}\right]d{\Omega}
	\label{phase field approximation for the fracture energy}
	\end{equation}

The variational approach for fracture \citep{francfort1998revisiting} states that the elastic energy drives the evolution of the phase field.  To ensure cracks only under tension, the elastic energy is decomposed into tensile and compressive parts \citep{borden2012phase}. We follow the decomposition of \citet{miehe2010phase} where the strain tensor $\bm\varepsilon$  is decomposed as follows:
	\begin{equation}
	  \left\{
	   \begin{aligned}
	\bm\varepsilon^+=\sum_{a=1}^d \langle\varepsilon_a\rangle^+\bm n_a\otimes\bm n_a \\ \bm\varepsilon^-=\sum_{a=1}^d \langle \varepsilon_a\rangle^-\bm n_a\otimes\bm n_a
	   \end{aligned}\right.
	\end{equation}

\noindent where $\bm\varepsilon^+$  and $\bm\varepsilon^-$  are the tensile and compressive strain tensors, respectively; $\varepsilon_a$  and  $\bm n_a$ are the principal strains and their directions. The operators $\langle\centerdot\rangle^+$  and $\langle\centerdot\rangle^-$  are defined as  $\langle\centerdot\rangle^+=(\centerdot+|\centerdot|)/2$, and $\langle\centerdot\rangle^-=(\centerdot-|\centerdot|)/2$. Using the decomposed strain tensor, the tensile and compressive parts of the elastic energy density are given by
	\begin{equation}
	  \left\{
	   \begin{aligned}
	\psi_{\varepsilon}^+(\bm \varepsilon) = \frac{\lambda}{2}\langle tr(\bm\varepsilon)\rangle^{+2}+\mu tr \left(\bm\varepsilon^{+2}\right) 
	\\ \psi_{\varepsilon}^-(\bm \varepsilon) = \frac{\lambda}{2}\langle tr(\bm\varepsilon)\rangle^{-2}+\mu tr \left(\bm\varepsilon^{-2}\right) 
	   \end{aligned}\right.
	\end{equation}

We follow \citet{borden2012phase} and assume that the phase field affects only the tensile part of the elastic energy density. Next, the stiffness reduction is modeled by the following equation:
	\begin{equation}
	\psi_{\varepsilon}(\bm\varepsilon)=\left[(1-k)(1-\phi)^2+k\right]\psi_{\varepsilon}^+(\bm \varepsilon)+\psi_{\varepsilon}^-(\bm \varepsilon)
	\label{decomposition of the elastic energy}
	\end{equation}

\noindent where $0<k\ll1$ is a model parameter that prevents the tensile part of the elastic energy density from disappearing and avoids  numerical singularity when the phase field $\phi$  tends to 1.

\subsection{Governing equations for evolution of the phase field}

Based on the phase field approximation for the fracture energy \eqref{phase field approximation for the fracture energy} and decomposition of the elastic energy \eqref{decomposition of the elastic energy}, the energy functional \eqref{functional2} can be rewritten as
	\begin{multline}
	L=\int_{\Omega}\left\{\left[(1-k)(1-\phi)^2+k\right]\psi_{\varepsilon}^+(\bm \varepsilon)+\psi_{\varepsilon}^-(\bm \varepsilon)\right\}d{\Omega}-\int_{\Omega}\alpha p \cdot (\nabla \cdot \bm u) d{\Omega}+\int_{\Omega}G_c\left[\frac{\phi^2}{2l_0}+\frac{l_0}2\frac{\partial\phi}{\partial x_i}\frac{\partial\phi}{\partial x_i}\right]d{\Omega}-\\ \int_{\Omega} b_iu_id{\Omega}-\int_{\partial\Omega_{h_i}} f_iu_idS
	\label{final functional}
	\end{multline}

The variational approach \citep{francfort1998revisiting} stresses that initiation, propagation and branching of the crack $\Gamma(\bm x,t)$ at  time $t\in[0,T]$ for ${\bm x} \in\Omega$ occur when the potential achieves a minimum value. Hence, we calculate the first variation of the functional $L$ and set it zero, yielding
	\begin{multline}
	\delta L=\underbrace{\int_{\partial \Omega_{h_i}}\left[(\sigma_{ij}-\alpha p\delta_{ij})m_j-f_i \right] \delta u_i d{S}}_{\textcircled{1}}
-\underbrace{\int_{\Omega}\left[(\sigma_{ij}-\alpha p \delta_{ij})_{,j}+b_i\right]\delta  u_i d{\Omega}}_{\textcircled{2}}
-\\ \underbrace{\int_{\Omega} \left[ 2(\phi-1)(1-k)\psi_{\varepsilon}^+ + \frac {G_c \phi}{l_0}-G_c l_0\frac{\partial^2\phi}{\partial x_i^2} \right]\delta\phi d{\Omega}}_{\textcircled{3}} + \underbrace{\int_{\partial\Omega}\left( \frac{\phi}{x_i}m_i\right)\delta\phi dS}_{\textcircled{4}}=0
	\label{first variation of the functional}
	\end{multline}

\noindent where $m_j$ is the component of the outward-pointing normal vector of the boundary and $\sigma_{ij}$ is the component of the effective trial stress tensor $\bm \sigma(\bm\varepsilon)$ given by:
	\begin{equation}
	\sigma_{ij}=\left [(1-k)(1-\phi)^2+k \right]\frac {\partial{\psi_\varepsilon^+}}{\partial {\varepsilon_{ij}}}+\frac {\partial{\psi_\varepsilon^-}}{\partial {\varepsilon_{ij}}}
	\end{equation}

\begin{equation}
	\bm \sigma=\left [(1-k)(1-\phi)^2+k \right]\left[\lambda \langle tr(\bm\varepsilon)\rangle^+ \bm I+ 2\mu \bm\varepsilon^+ \right]+\lambda \langle tr(\bm\varepsilon)\rangle^- \bm I+ 2\mu \bm\varepsilon^-
	\end{equation}

\noindent where $\bm I$ is the identity tensor $\in \mathbb R^{d\times d}$.

We recall the Cauchy stress tensor $\bm\sigma^{por}$:
	\begin{equation}
	\bm \sigma^{por}(\bm\varepsilon)=\bm \sigma(\bm\varepsilon)-\alpha p \bm I,\hspace{0.5cm} in \hspace{0.1cm} \Omega\times(0,T]
	\end{equation}

Because of the arbitrariness of the variation $\delta u$ and $\delta \phi$, Eq. \eqref{first variation of the functional}$\textcircled{2}$ and $\textcircled{3}$ give rise to the governing equations:
	\begin{equation}
	  \left\{
	   \begin{aligned}
	\frac {\partial {\sigma_{ij}^{por}}}{\partial x_j}+b_i=0, &\hspace{0.5cm} in \hspace{0.1cm} \Omega\times(0,T]
	\\ \left[\frac{2l_0(1-k)\psi_{\varepsilon}^+}{G_c}+1\right]\phi-l_0^2\frac{\partial^2 \phi}{\partial {x_i^2}}=\frac{2l_0(1-k)\psi_{\varepsilon}^+}{G_c}, &\hspace{0.5cm} in \hspace{0.1cm} \Omega\times(0,T]
	   \end{aligned}\right.
	\label{governing equations 0}
	\end{equation}

The variational approach \citep{francfort1998revisiting} also emphasizes that the irreversibility condition $\Gamma(\bm x,s)\in\Gamma(\bm x,t)(s<t)$ must be satisfied, which ensures that a crack cannot heal. To ensure a monotonically increasing phase field, the irreversibility condition during compression or unloading is accomplished by introducing a strain-history field  $H(\bm x,t)$ \citep{miehe2010phase,miehe2010thermodynamically,borden2012phase} defined by
	\begin{equation}
	H(\bm x,t) = \max \limits_{s\in[0,t]}\psi_\varepsilon^+\left(\bm\varepsilon(\bm x,s)\right), \hspace{0.5cm} in \hspace{0.1cm} \Omega\times(0,T]
	\end{equation}

The history field $H$ satisfies the Kuhn-Tucker conditions for loading and unloading
	\begin{equation}
	\psi_\varepsilon^+-H\le0,\hspace{0.5cm}\dot{H}\ge0,\hspace{0.5cm}\dot{H}(\psi_\varepsilon^+-H)=0
	\end{equation}
	More details can be found in \citep{miehe2010phase}. Replacing $\psi_\varepsilon^+$  by  $H(\bm x,t)$  in Eq. \eqref{governing equations 0}, the strong form is obtained by
	\begin{equation}
	  \left\{
	   \begin{aligned}
	\frac {\partial {\sigma_{ij}^{por}}}{\partial x_i}+b_i=0, &\hspace{0.5cm} in \hspace{0.1cm} \Omega\times(0,T]
		\\ \left[\frac{2l_0(1-k)H}{G_c}+1\right]\phi-l_0^2\frac{\partial^2 \phi}{\partial {x_i^2}}=\frac{2l_0(1-k)H}{G_c}, &\hspace{0.5cm} in \hspace{0.1cm} \Omega\times(0,T]
	\label{governing equation1}
	   \end{aligned}\right.
	\end{equation}

In addition, Eq. \eqref{first variation of the functional}$\textcircled{1}$ and $\textcircled{4}$ give rise to the Neumann conditions of the displacement and phase field,
	\begin{equation}
	  \left\{
	   \begin{aligned}
	&\sigma_{ij}^{por}m_j=f_i, \hspace{1cm} &on\hspace{0.5cm} \partial\Omega_{h_i}\times(0,T]
	\\ &\frac{\partial \phi}{\partial x_i} m_i = 0, &on\hspace{0.5cm} \partial\Omega\times(0,T]
	\label{boundary condition of the phase field}
	\end{aligned}\right.
	\end{equation}

\subsection{Governing equations for the flow field}

To formulate the flow equations in the domain $\Omega$, we divide the whole domain into three parts: $\Omega_r(t)$, $\Omega_f(t)$ and $\Omega_t(t)$. $\Omega_r(t)$ represents the unbroken domain (reservoir domain) and $\Omega_f(t)$ is the fracture domain. $\Omega_t(t)$ is the transition domain betwen $\Omega_r(t)$ and $\Omega_f(t)$. Here, we follow \citet{lee2016pressure} and use the phase field as an indicator function. Two thresholds $c_1$ and $c_2$ are set to separate the three flow domains. We consider a subdomain as the fracture domain $\Omega_f(t)$ if $\phi\ge c_2$ and as the reservoir domain $\Omega_r(t)$ if $\phi\le c_1$. In the transition domain, $c_1<\phi<c_2$.

We use Darcy's law to describe the flow field in the poroelastic medium. For the reservoir domain $\Omega_r(t)$, the mass conservation leads to 
	\begin{equation}
	\frac{\partial}{\partial t}(\varepsilon_{pr}\rho_r)+\nabla\cdot(\rho_r\bm v_r)=q_r-\rho_r\alpha_r\frac{\partial \varepsilon_{vol}}{\partial t}
	\label{mass conservation of the reservoir domain}
	\end{equation}  

\noindent where $\rho_r$, $q_r$, $\varepsilon_{pr}$, and $\alpha_r$ are the density of fluid, source term, porosity, and Biot coefficient in the reservoir domain, respectively; $\varepsilon_{vol}=\nabla\cdot\bm u$ is the volumetric strain of $\Omega_r(t)$.

Darcy's velocity in $\Omega_r(t)$ is given by
	\begin{equation}
	\bm v_r=-\frac{K_r}{\mu_r}(\nabla p+\rho_r\bm g)
	\label{velocity of the reservoir domain}
	\end{equation} 

\noindent where $K_r$ and $\mu_r$ are the permeability and fluid viscosity of $\Omega_r(t)$, respectively; $\bm g$ is the gravity.

Based on the storage model in \citep{biot1962mechanics}, we have 
	\begin{equation}
	\frac{\partial}{\partial t}(\varepsilon_{pr}\rho_r) = \rho_r S_r \frac{\partial p}{\partial t}
	\end{equation}

\noindent where $S_r$, the storage coefficient of $\Omega_r$, is given by
 	\begin{equation}
	S_r=\varepsilon_{pr}c_r+\frac{(\alpha_r-\varepsilon_{pr})(1-\alpha_r)}{K_{Vr}}
	\end{equation}

\noindent with $c_r$ the fluid compressibility and $K_{Vr}$ the bulk modulus of the reservoir domain. Thus, the equation of mass conservation \eqref{mass conservation of the reservoir domain} translates to
	\begin{equation}
	\rho_r S_r \frac{\partial p}{\partial t}+\nabla\cdot(\rho_r\bm v_r)=q_r-\rho_r\alpha_r\frac{\partial \varepsilon_{vol}}{\partial t}
	\label{governing equation of the reservoir domain}
	\end{equation}  

For the fracture domain $\Omega_f(t)$, the volumetric strain vanishes from the equation of mass conservation:
	\begin{equation}
	\rho_f S_f \frac{\partial p}{\partial t}+\nabla\cdot(\rho_f\bm v_f)=q_f
	\label{governing equation of the fracture domain}
	\end{equation} 

\noindent where $\rho_f$, $S_f$, and $q_f$ are the fluid density, storage coefficient, and source term in the fracture domain $\Omega_f(t)$. 

The storage coefficient $S_f$ is equal to the fluid compressibility $c_f$ and the Darcy's velocity in $\Omega_f(t)$ is given by 
	\begin{equation}
	\bm v_f=-\frac{K_f}{\mu_f}(\nabla p+\rho_f\bm g)
	\label{velocity of the fracture domain}
	\end{equation} 

\noindent with $K_f$ and $\mu_f$ the permeability and fluid viscosity of $\Omega_f(t)$. 

We follow \citet{lee2016pressure} and define two linear indicator functions to connect the transition domain with the reservoir and fracture domains: $\chi_r$ and $\chi_f$, which satisfies
	\begin{equation}
	\chi_r(\cdot,\phi):= \chi_r(\bm x, t,\phi)=1 \quad in \quad \Omega_r(t),\quad and \quad \chi_r(\cdot,\phi)= 0 \quad in \quad \Omega_f(t)
	\end{equation} 
	\begin{equation}
		\chi_f(\cdot,\phi):= \chi_f(\bm x, t,\phi)=1 \quad in \quad \Omega_f(t),\quad and \quad \chi_f(\cdot,\phi)= 0 \quad in \quad \Omega_r(t)
		\end{equation} 

In the transition domain, the two linear functions are defined as
	\begin{equation}
	\chi_r(\cdot,\phi)=\frac{c_2-\phi}{c_2-c_1},\quad and \quad \chi_f(\cdot,\phi)=\frac{\phi-c_1}{c_2-c_1}
	\label{functions}
	\end{equation} 

\begin{figure}[htbp]
	\centering
	\subfigure[]{\includegraphics[height = 5cm]{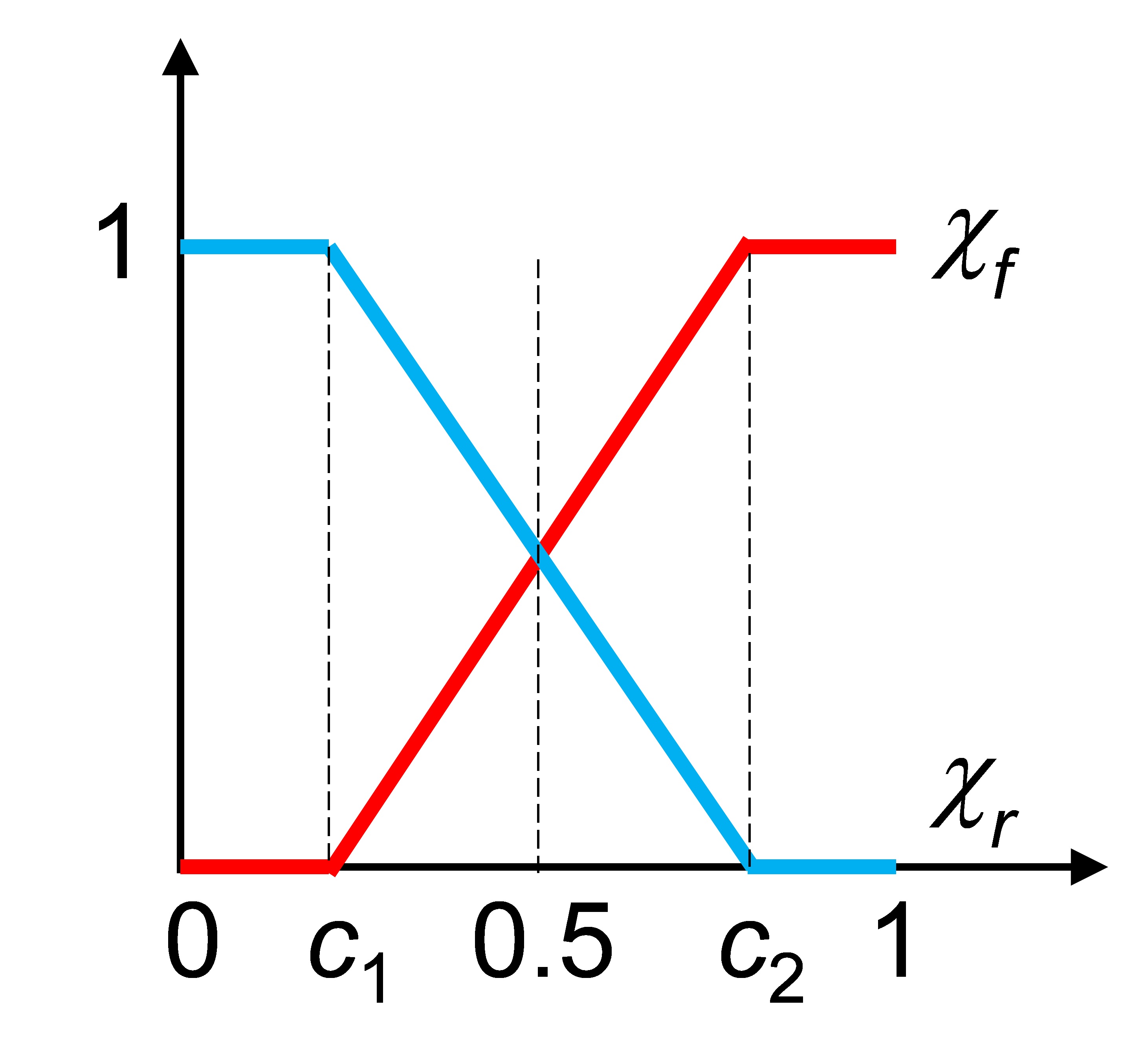}}\hspace{1cm}
	\subfigure[]{\includegraphics[height = 5cm]{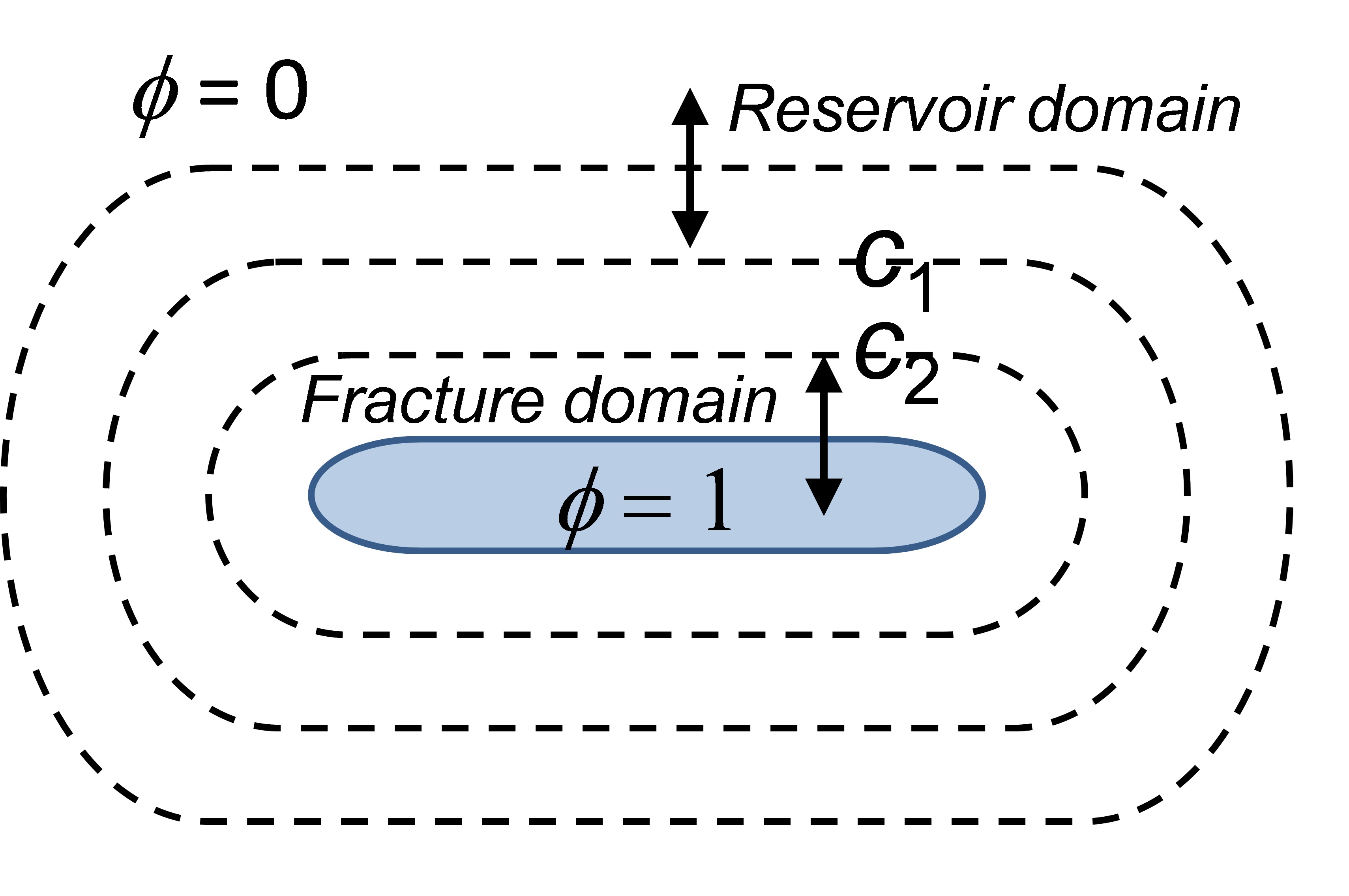}}
	\caption{(a) Linear indicator functions $\chi_r$ and $\chi_f$ and (b) the reservoir and fracture domains}
	\label{(a) Linear indicator functions and (b) the reservoir and fracture domains}
	\end{figure}

Figure \ref{(a) Linear indicator functions and (b) the reservoir and fracture domains} shows more details of the linear indicator functions $\chi_r$ and $\chi_f$ as well as illustration of the fracture and reservoir domains. The fluid and solid properties of the transition domain $\Omega_t$ are obtained by interpolation of the reservoir and fracture domains with the indicator functions $\chi_r$ and $\chi_f$. Thus, the mass conservation in the transition domain is rewritten as
	\begin{equation}
	\rho S \frac{\partial p}{\partial t}+\nabla\cdot(\rho\bm v)=q_m-\rho\alpha\chi_r\frac{\partial \varepsilon_{vol}}{\partial t}
	\label{mass conservation of the whole domain}
	\end{equation}  

\noindent  where $\rho=\rho_r\chi_r+\rho_f\chi_f$, $\alpha=\alpha_r\chi_r+\alpha_f\chi_f$, $q_m$ is the source term, and the storage coefficient $S$ is represented by
 	\begin{equation}
	S=\varepsilon_pc+\frac{(\alpha-\varepsilon_p)(1-\alpha)}{K_{Vr}}
	\end{equation}

\noindent with $c=c_r\chi_r+c_f\chi_f$. Note that $\varepsilon_p=0$ and $\alpha=1$ for the fracture domain and thereby $\varepsilon_p=\varepsilon_{pr}\chi_r$ and $\alpha=\alpha_{r}\chi_r+\chi_f$. 

Eq. \eqref{mass conservation of the whole domain} is also applicable for the whole domain $\Omega$ because it can be transformed into Eq. \eqref{governing equation of the reservoir domain} for $\Omega_r$ and Eq. \eqref{governing equation of the fracture domain} for $\Omega_f$. The Darcy's velocity $\bm v$ is then calculated by
	\begin{equation}
	\bm v=-\frac{K}{\mu}(\nabla p+\rho\bm g)
	\label{velocity of the whole domain}
	\end{equation} 

\noindent where $K=K_{r}\chi_r+K_f\chi_f$ is the effective permeability and $\mu=\mu_{r}\chi_r+\mu_f \chi_f$ is the effective fluid viscosity. Finally, the governing equation for the flow field in the porous domain $\Omega$ is rewritten in terms of the fluid pressure $p$:
	\begin{equation}
	\rho S \frac{\partial p}{\partial t}-\nabla\cdot \frac{\rho K}{\mu}(\nabla p+\rho\bm g)=q_m-\rho\alpha\chi_r\frac{\partial \varepsilon_{vol}}{\partial t}
	\label{governing equation of the whole domain}
	\end{equation}

Note that the determination of the coefficients in the phase-field model will be tackled in future research. Within Comsol, it is easily possible to  include an indirect coupling such as the dependence of the permeability of the fracture domain on the crack opening. Relevant contributions to such an indirect coupling are for instance proposed in \citep{yeo1998effect, alm1999hydro, mcdermott2004hydraulic, cammarata2007hydro}. In this context, we also mention the experimental work by \citet{yeo1998effect} who explored the effect of shear deformation on the aperture and permeability of rock fracture. However, the crack opening cannot be extracted in a straight forward manner in the PFM because the sharp crack is represented by a smooth phase field. Therefore, for simplicity, we use an unchanged fluid property in the fracture domain in our presented examples, which also shows favorable results.

\subsection{Initial and boundary conditions}

For time-transient problems, the following initial conditions must be fulfilled:
	\begin{equation}
	  \left\{
	   \begin{aligned}
	&\bm u(\bm x,0)=\bm u_0(\bm x)\hspace{2cm} &\bm x\in\Omega
	\\ &p(\bm x,0)=p_0(\bm x)\hspace{2cm} &\bm x\in\Omega
	\\&\phi(\bm x,0)=\phi_0(\bm x)\hspace{2cm} &\bm x\in\Omega
	\end{aligned}\right.
	\label{Initial condition}
	\end{equation}

Here, the initial phase field  $\phi_0=1$ in a local domain can be used to model a pre-existing crack \citep{borden2012phase}. Boundary conditions for the displacement and phase field are introduced in Subsections 2.1 and 2.3. For the fluid pressure, the Dirichlet  condition on $\partial\Omega_D$ and Neumann condition on $\partial\Omega_N$ with $\partial\Omega_D\cap \partial\Omega_N =\emptyset$ are prescribed:
	\begin{equation}
	p=p_D \hspace{2cm}on\quad\partial\Omega_D\times(0,T]
	\end{equation}
	\begin{equation}
	-\bm m\cdot \rho\bm v=M_N  \hspace{2cm}on\quad\partial\Omega_N\times(0,T]
	\end{equation}

\noindent with $p_D$ the prescribed pressure on the Dirichlet boundary and $M_N$ the mass flux on the Newman boundary.

\section {Numerical implementation}
\subsection{Finite element discretization}\label{Finite element discretization}
The proposed phase field  approach is applied in the framework of finite element method. Thus, we first derive the weak form of all the governing equations as
	\begin{equation}
	\int_{\Omega}\left[-(\bm\sigma-\alpha p\bm I):\delta \bm {\varepsilon}\right] \mathrm{d}\Omega +\int_{\Omega}\bm b \cdot \delta \bm u \mathrm{d}\Omega +\int_{\Omega_{h_i}}\bm f \cdot \delta \bm u \mathrm{d}S=0
	\label{weak form 1}
	\end{equation}
\noindent ,
	\begin{equation}
	\int_{\Omega}-2(1-k)H(1-\phi)\delta\phi\mathrm{d}\Omega+\int_{\Omega}G_c\left(l_0\nabla\phi\cdot\nabla\delta\phi+\frac{1}{l_0}\phi\delta\phi\right)\mathrm{d}\Omega=0
	\label{weak form 2}
	\end{equation}
\noindent and
	\begin{equation}
	\int_{\Omega} \rho S \frac{\partial p}{\partial t}\delta p\mathrm{d}\Omega-\int_{\Omega} \rho \bm v \cdot\nabla\delta p \mathrm{d}\Omega=\int_{\partial\Omega}M_n\mathrm{d}S+\int_{\Omega}\left(q_m-\rho\alpha\chi_r\frac{\partial \varepsilon_{vol}}{\partial t}\right)\mathrm{d}\Omega
	\label{weak form 3}
	\end{equation}  

The nodal values for the three fields ($\bm u$, $\phi$, and $p$) are defined as $\bm u_i$, $\phi_i$, and $p_i$. Afterwards, the fields are discretized by the standard 3D notation as
	\begin{equation}
	\bm u = \sum_i^{n}N_i \bm u_i,\hspace{0.5cm} \phi = \sum_i^{n}N_i \phi_i,\hspace{0.5cm} p = \sum_i^{n}N_i p_i
	\end{equation}
\noindent where $n$ is the total number of nodes in each element and $N_i$ is the shape function of node $i$. The gradients of the three fields are thereby given by
	\begin{equation}
	\bm \varepsilon = \sum_i^{n}\bm B_i^u \bm u_i,\hspace{0.5cm} \nabla\phi = \sum_i^{n}\bm B_i^{\phi} \phi_i,\hspace{0.5cm} \nabla p = \sum_i^{n}\bm B_i^{p} p_i
	\end{equation}
\noindent where $\bm B_i^u$, $\bm B_i^\phi$, and $\bm B_i^p$ are derivatives of the shape functions:
	\begin{equation}
	\bm B_i^u=\left[
	\begin{array}{ccc}
	N_{i,x}&0&0\\
	0&N_{i,y}&0\\
	0&0&N_{i,z}\\
	N_{i,y}&N_{i,x}&0\\
	0&N_{i,z}&N_{i,y}\\
	N_{i,z}&0&N_{i,x}
	\end{array}\right],
	\hspace{0.5cm}\bm B_i^{\phi}=\bm B_i^p=\left[
	\begin{array}{ccc}
	N_{i,x}\\
	N_{i,y}\\
	N_{i,z}
	\end{array}\right]\label{gradient of shape function 3D}
	\end{equation}

For 2D, the above equations \eqref{gradient of shape function 3D} can be simplified after a little adjustment. Due to the arbitrariness of the test functions, the external force $\bm F_i^{u,ext}$ and inner force $\bm F_i^{u,int}$ of the mechanical field are described by
	\begin{equation}
	\left\{
	\begin{aligned}
	\bm F_i^{u,ext} &= \int_{\Omega}N_i\bm b \mathrm{d}\Omega+ \int_{\Omega_{h_i}}N_i\bm f \mathrm{d}S+ \int_{\Omega}[\bm B_{i}^{u}]^{\mathrm T}\alpha p \bm I \mathrm{d}\Omega\\
	\bm F_i^{u,int} &= \int_{\Omega}[\bm B_i^u]^{\mathrm T}\bm\sigma \mathrm{d}\Omega
	\end{aligned}
	\right.
	\end{equation}

The inner force term of the phase field are also obtained by
	\begin{equation} 
	F_i^{\phi,int} = \int_{\Omega} -2(1-k)(1-\phi)H N_i+G_c\left(l_0[\bm B_i^{\phi}]^{\mathrm T}\nabla \phi+\frac 1 {l_0}\phi N_i \right ) \mathrm{d}\Omega
	\end{equation}

Finally, when the gravity is neglected, the inner force $ F_i^{p,int}$, viscous force $F_i^{p,vis}$, and external force $F_i^{p,ext}$ of the pressure field is given by
	\begin{equation}
	\left\{
	\begin{aligned}
	F_i^{p,int} &= \int_{\Omega}[\bm B_i^p]^{\mathrm T}\frac{\rho K}{\mu}\nabla p \mathrm{d}\Omega\\
	F_i^{p,vis} &= \int_{\Omega}N_i \rho S \frac{\partial p}{\partial t} \mathrm{d}\Omega\\
	F_i^{p,ext} &=  \int_{\Omega}N_i \left( q_m-\rho\alpha\chi_r\frac{\partial \varepsilon_{vol}}{\partial t} \right) \mathrm{d}\Omega+ \int_{\partial\Omega_{N}}N_i M_N \mathrm{d}S
	\end{aligned}
	\right.
	\end{equation}

Thus, according to Eqs. \eqref{weak form 1}, \eqref{weak form 2} and \eqref{weak form 3}, contribution of node $i$ to the residual of the discrete equations of stress equilibrium, evolution of phase field, and mass conservation of pressure field is given as
	\begin{equation} 
	\left\{
	\begin{aligned}
	\bm R_i^{u}&=\bm F_i^{u,ext}-\bm F_i^{u,int}\\
	R_i^{\phi}&=- F_i^{\phi,int}\\
	R_i^{p}&=F_i^{p,ext}- F_i^{p,int}- F_i^{p,vis}
	\end{aligned}
	\right.
	\end{equation}

We use the segregated scheme to solve the displacement, phase field and fluid pressure. The segregated scheme ensures that the energy functional of each field remains convex. The Newton-Raphson approach is adopted to obtain $\bm R_i^{u}=0$, $ R_i^{\phi}=0$, and $ R_i^{p}=0$, respectively. The tangents on the element level are thereby calculated by
		\begin{equation}
		\left\{
		\begin{aligned} 
		\bm K_{ij}^{uu}&=\frac{\partial \bm F_i^{u,int}}{\partial \bm u_j}=\int_{\Omega}[\bm B_i^u]^{\mathrm T}\bm D [\bm B_j^u] \mathrm{d} \Omega\\
		\bm K_{ij}^{\phi\phi}&=\frac{\partial F_i^{\phi,int}}{\partial \phi_j}=\int_{\Omega}\left\{[\bm B_i^\phi]^{\mathrm T} G_cl_0 [\bm B_j^\phi]+N_i\left(2(1-k)H+\frac{G_c}{l_0}\right)N_j\right\} \mathrm{d} \Omega\\
		\bm K_{ij}^{pp}&=\frac{\partial F_i^{p,int}}{\partial p_j}=\int_{\Omega}[\bm B_i^p]^{\mathrm T} \frac{\rho K}{\mu} [\bm B_j^p] \mathrm{d} \Omega
		\end{aligned}
		\right.
		\end{equation}

\noindent where $\bm D$ is the elasticity tensor of four order given by $\bm D = {\partial \bm\sigma}/{\partial \bm\varepsilon}$.

\subsection{Comsol implementation}
The phase field model is implemented into Comsol. We establish three main modules first: Poroelasticity Module, History-strain Module and Phase Field Module.  The Poroelasticity Module is used to solve the displacement field and fluid pressure. The History-strain Module and Phase Field Module are established to solve the other two fields $H$ and $\phi$. These modules are all written in strong forms and solved based on standard finite element discretization in space domain and finite difference discretization in time domain. The implementation in Comsol needs many intermediate variables, such as the elastic energy, principal strain and direction vectors, and thereby we also establish a pre-set Storage Module to evaluate and store these intermediate variables during each time step.

\subsection{Module setup}
The Poroelasticity Module is set up based on a linear elastic material library and a transient formulation of Darcy's law. The boundary and initial conditions in the Poroelasticity Module are implemented as shown in Section 2. However, the elasticity matrix in a time step requires modification. The elasticity matrix comes from the elasticity tensor of fourth order $\bm D$ given by
	\begin{equation}\begin{aligned}
	\bm D = \frac {\partial \bm\sigma}{\partial \bm\varepsilon}=\lambda\left\{ \left[(1-k)(1-\phi)^2+k \right]H_\varepsilon(tr(\bm\varepsilon))+H_\varepsilon(-tr(\bm\varepsilon))\right\}\bm J +\\
2\mu\left\{\left[(1-k)(1-\phi)^2+k \right]\frac{\partial \bm\varepsilon_+}{\partial \bm\varepsilon}+\frac{\partial \bm\varepsilon_-}{\partial \bm\varepsilon}\right\}
	\end{aligned}
	\end{equation}

\noindent where $H_\varepsilon \langle x \rangle$  is a Heaviside function: $H_\varepsilon \langle x \rangle=1$  if  $x>0$ and $H_\varepsilon \langle x \rangle=0$  if $x\leq 0$, and $J_{ijkl}=\delta_{ij}\delta_{kl}$  with $\delta_{ij}$ and $\delta_{kl}$ the Kronecker symbols. Finally, the stiffness matrix $\bm G$ is rewritten in a standard material data ordering:
	\begin{equation}
	\bm G= \left [
	\begin{array}{cccccc}
	G_{1111} & G_{1122} & G_{1133} & G_{1112} & G_{1123} & G_{1113}\\
	G_{2211} & G_{2222} & G_{2233} & G_{2212} & G_{2223} & G_{2213}\\
	G_{3311} & G_{3322} & G_{3333} & G_{3312} & G_{3323} & G_{3313}\\
	G_{1211} & G_{1222} & G_{1233} & G_{1212} & G_{1223} & G_{1213}\\
	G_{2311} & G_{2322} & G_{2333} & G_{2312} & G_{2323} & G_{2313}\\
	G_{1311} & G_{1322} & G_{1333} & G_{1312} & G_{1323} & G_{1313}
	\end{array}
	\right ]
	\end{equation}

\noindent with $G_{ijkl}=\bar G_{ijkl}+\tilde G_{ijkl}$.

 $\bar G_{ijkl}$ is related to trace of the strain $tr(\bm \varepsilon)$:
	\begin{equation}
	\bar G_{ijkl}=\lambda\left\{ \left[(1-k)(1-\phi)^2+k \right]H_\varepsilon(tr(\bm\varepsilon))+H_\varepsilon(-tr(\bm\varepsilon))\right\} \delta_{ij}\delta_{kl}
	\end{equation}

Based on the algorithm for fourth-order isotropic tensor \citep{miehe1998comparison}, $\tilde G_{ijkl}$ is expressed as
	\begin{equation}
	\tilde G_{ijkl}=2\mu\left\{ \left[(1-k)(1-\phi)^2+k \right]P_{ijkl}^++P_{ijkl}^-\right\}
	\label{Pijkl}
	\end{equation}

\noindent where	
	\begin{equation}
	P_{ijkl}^+ = \sum_{a=1}^3\sum_{b=1}^3 H_\varepsilon(\varepsilon_a)\delta_{ab}n_{ai}n_{aj}n_{bk}n_{bl}+\sum_{a=1}^3\sum_{b\neq a}^3 \frac 1 2 \frac {\langle \varepsilon_a\rangle_+ - \langle \varepsilon_b\rangle_+}{\varepsilon_a-\varepsilon_b}n_{ai}n_{bj}(n_{ak}n_{bl}+n_{bk}n_{al})
	\label{P+ijkl}
	\end{equation}
\noindent and 
	\begin{equation}
	P_{ijkl}^- = \sum_{a=1}^3\sum_{b=1}^3 H_\varepsilon(-\varepsilon_a)\delta_{ab}n_{ai}n_{aj}n_{bk}n_{bl}+\sum_{a=1}^3\sum_{b\neq a}^3 \frac 1 2 \frac {\langle \varepsilon_a\rangle_- - \langle \varepsilon_b\rangle_-}{\varepsilon_a-\varepsilon_b}n_{ai}n_{bj}(n_{ak}n_{bl}+n_{bk}n_{al})
	\label{P-ijkl}
	\end{equation}

\noindent with $n_{ai}$ the $i$-th component of vector $\bm n_a$.

Because of the fractional form given in Eqs. \eqref{P+ijkl} and \eqref{P-ijkl},  Eq. \eqref{Pijkl} cannot be calculated if $\varepsilon_a=\varepsilon_b$. To better apply  Eq.\eqref{Pijkl} in Comsol, we therefore refer to \citep{miehe1993computation} and use a ``perturbation''  for the principal strains:
	\begin{equation}
	  \left\{
	   \begin{aligned}
	&\varepsilon_1 = \varepsilon_1(1+\delta)\hspace{0.5cm} &if\hspace{0.1cm}\varepsilon_1 = \varepsilon_2
	\\ &\varepsilon_3 = \varepsilon_3(1-\delta)\hspace{0.5cm} &if\hspace{0.1cm}\varepsilon_2 = \varepsilon_3
	\end{aligned}\right.
	\end{equation}

\noindent with the perturbation $\delta=1\times 10^{-9}$ for this paper. The second principal strain and volumetric strain remain unchanged. 

We establish the Phase Field Module by revising the Helmholtz equation in Comsol. Coefficients of the Helmholtz equation must have the same form as the governing equation \eqref{governing equation1}. The boundary condition in Eq. \eqref{boundary condition of the phase field} and initial condition \eqref{Initial condition} are also implemented in this module. 

Comsol has stable Distributed ODEs and DAEs Interface to solve distributed ordinary differential equations (ODE) and differential-algebraic equations (DAE). Thus, we use the Distributed ODEs and DAEs Interfaces to establish the History-strain Module, in which we do not really solve the history-strain field but call a ``previous solution" function of Comsol solver to record and update $H(\bm x, t)$. The format of the equations written into the History-strain Module is as follows,
	\begin{equation}
	H =  \left\{
	   \begin{aligned}
	&H,\hspace{2cm} &H\ge \psi_{\varepsilon}^+
	\\&\psi_{\varepsilon}^+,&H< \psi_{\varepsilon}^+
	\end{aligned}\right.
	\end{equation}

The History-strain Module also needs initial conditions, commonly, $H_0(\bm x)=0$. However, pre-existing cracks can be generated by introducing the following initial conditions \citep{borden2012phase}:
	\begin{equation}
	H_0(\bm x) =  \left\{
	   \begin{aligned}
	&\frac {BG_c}{2l_0}\left [1-\frac {2d(\bm x,l)}{l_0}\right ],\hspace{0.1cm}&d(\bm x, l)\leq \frac {l_0} 2
	\\&0,\hspace{0.5cm} & d(\bm x, l)> \frac {l_0} 2
	\end{aligned}\right.
	\end{equation}

When $d = 0$, substituting $H_0$ into \eqref{governing equation1}$_2$ will give rise to
	\begin{equation}
	B=\frac\phi {(1-k)(1-\phi)}
	\end{equation}

If $\phi$ becomes quite close to 1, $B$  will become quite large. Therefore, we follow \citep{borden2012phase} and adopt $B=1\times 10^3$ to artificially generate the initial cracks if necessary.

\subsection{Segregated scheme}

\begin{figure}[htbp]
	\centering
	\includegraphics[width = 10cm]{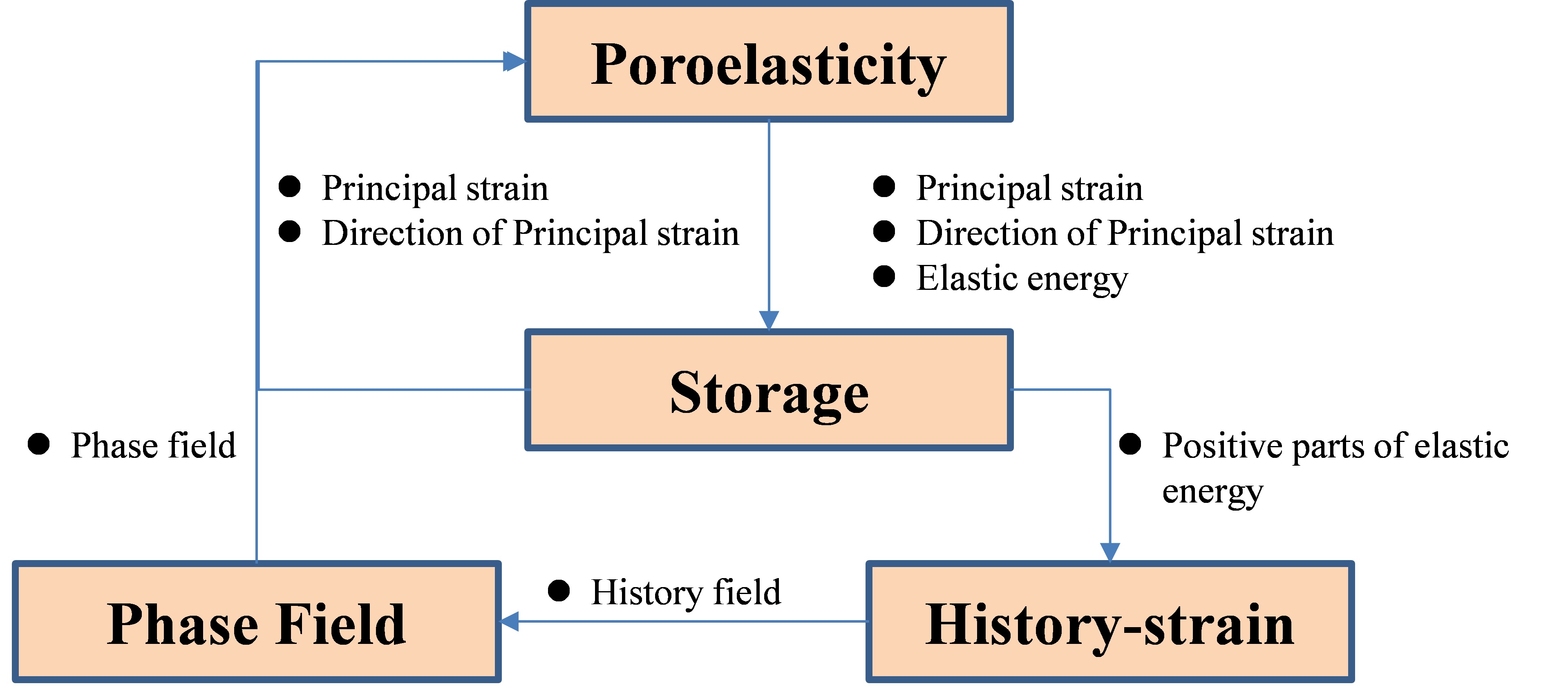}
	\caption{Relationship between the established modules}
	\label{Relationship between the established modules}
	\end{figure}

\begin{figure}[htbp]
	\centering
	\includegraphics[width = 15cm]{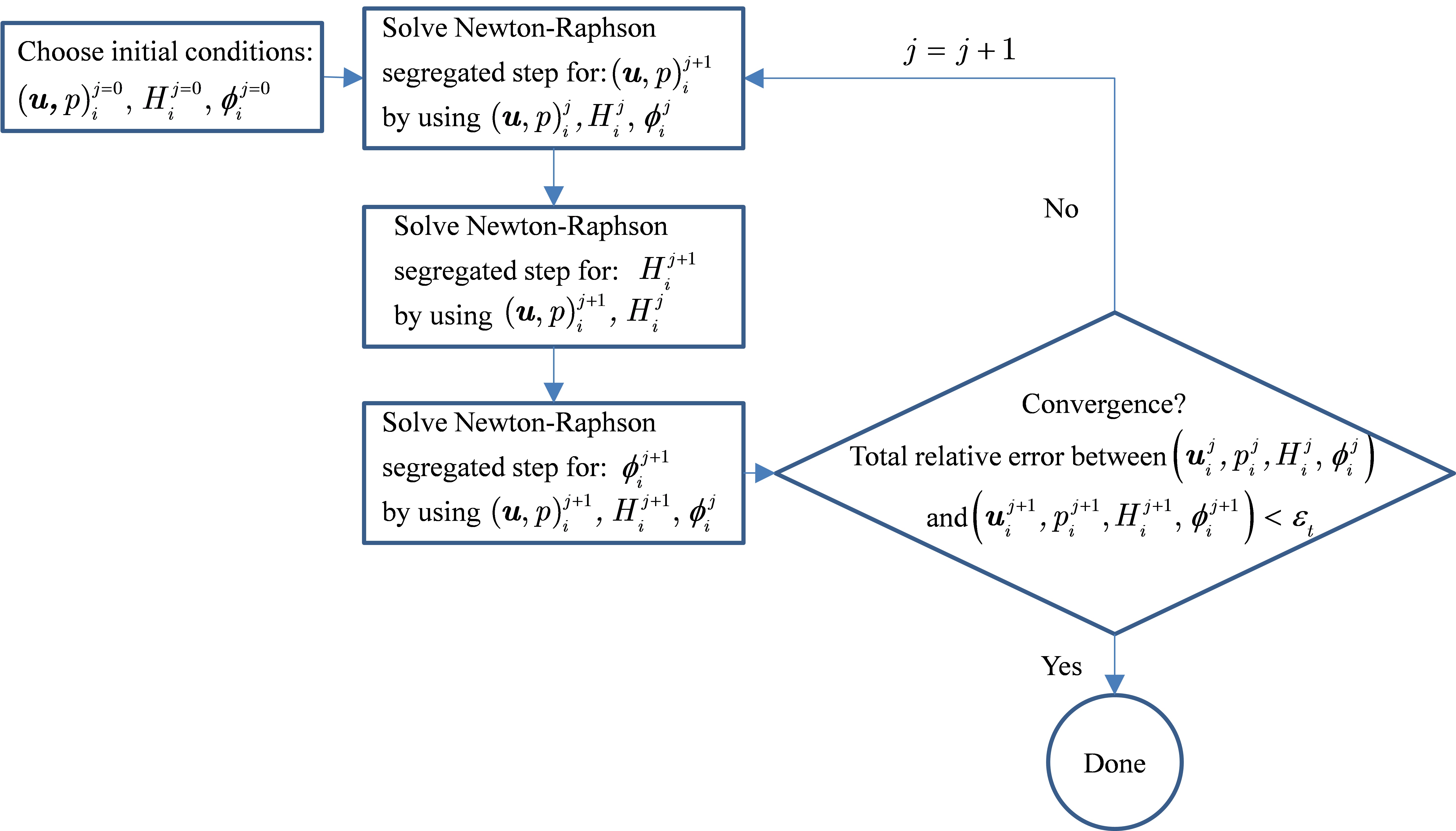}
	\caption{Segregated scheme for the coupled calculation in phase field modeling}
	\label{Segregated scheme for the coupled calculation in phase field modeling}
	\end{figure}

Figure \ref{Relationship between the established modules} shows all the established modules and their relationship. The Storage Module stores the principal strains and the direction of principal strain from the Poroelasticity Module. Some intermediate variables such as the elastic energy and the component $G_{ijkl}$ are also calculated in the Storage Module. The History-strain module then uses the positive part of the elastic energy $\psi_{\varepsilon}^+$ calculated from the Storage Module to solve and updated the local history strain field.  Afterwards, the Phase Field Module utilizes the updated history strain $H$ to solve the phase field. Because of the variation in the phase field, the mechanical and hydraulic properties of the domain need to be updated. Therefore, the stiffness matrix of the Poroelasticity Module is revised according to the previously stored intermediate variables and the updated phase field. The three types of domain $\Omega_R(t)$,  $\Omega_F(t)$, and  $\Omega_T(t)$  are also distinguished and assigned the updated hydraulic property. Finally, the displacement and pressure in the Poroelasticity Module are coupled to be solved. Because Fig. \ref{Relationship between the established modules} shows strong coupling of all the established modules, we use a segregated scheme to solve the coupled system of equations as indicated in Fig. \ref{Segregated scheme for the coupled calculation in phase field modeling}. The segregated scheme will make the simulation much easier to converge.

The segregated scheme means that all the fields are solved independently in one time step. We place the displacement and pressure in one segregated step and solve them together. Afterwards, the history strain and phase field are in another two segregated steps. Thus, the equations of poroelasticity (displacement and pressure), history strain and phase field are solved independently. The implicit Generalized-$\alpha$ method \citep{borden2012phase} is used to ensure unconditional stability for the calculation. When the time reaches $t_i$, the linear extrapolation of the previous solution provides the initial guess for the three segregated steps ($(\bm u,p)_i^{j=0}$, $H_i^{j=0}$ and $\phi_i^{j=0}$). Then, the Newton-Raphson method is used to solve each segregated step. For a new iteration step $j+1$ in the given time step $i$,  the displacement and pressure $(\bm u,p)_i^{j+1}$ is first solved  by using the results from previous iteration step $j$ ($(\bm u,p)_i^j$, $H_i^j$ and $\phi_i^j$) after a Newton-Raphson iteration. Based on the updated displacement $\bm u_i^{j+1}$, the history strain field is then updated. The Newton-Raphson iteration is subsequently applied to achieve a new phase field $\phi_i^{j+1}$ by $\bm u_i^{j+1}$, $H_i^{j+1}$. Finally,  the total relative error of the solutions from the previous and present iteration steps $\varepsilon_r$ is calculated. If $\varepsilon_r$ is less than the tolerance $\varepsilon_t$, the calculation for the time step $i$ is finished and the time will increase to $t_{i+1}$. If $\varepsilon_r\ge\varepsilon_t$, a new iteration step will be used until $\varepsilon_r<\varepsilon_t$. The tolerance $\varepsilon_t=1\times 10^{-4}$ is chosen in this paper. Thus, we will obtain all the solutions at all time points  by the segregated scheme.

\begin{figure}[htbp]
	\centering
	\includegraphics[width = 12cm]{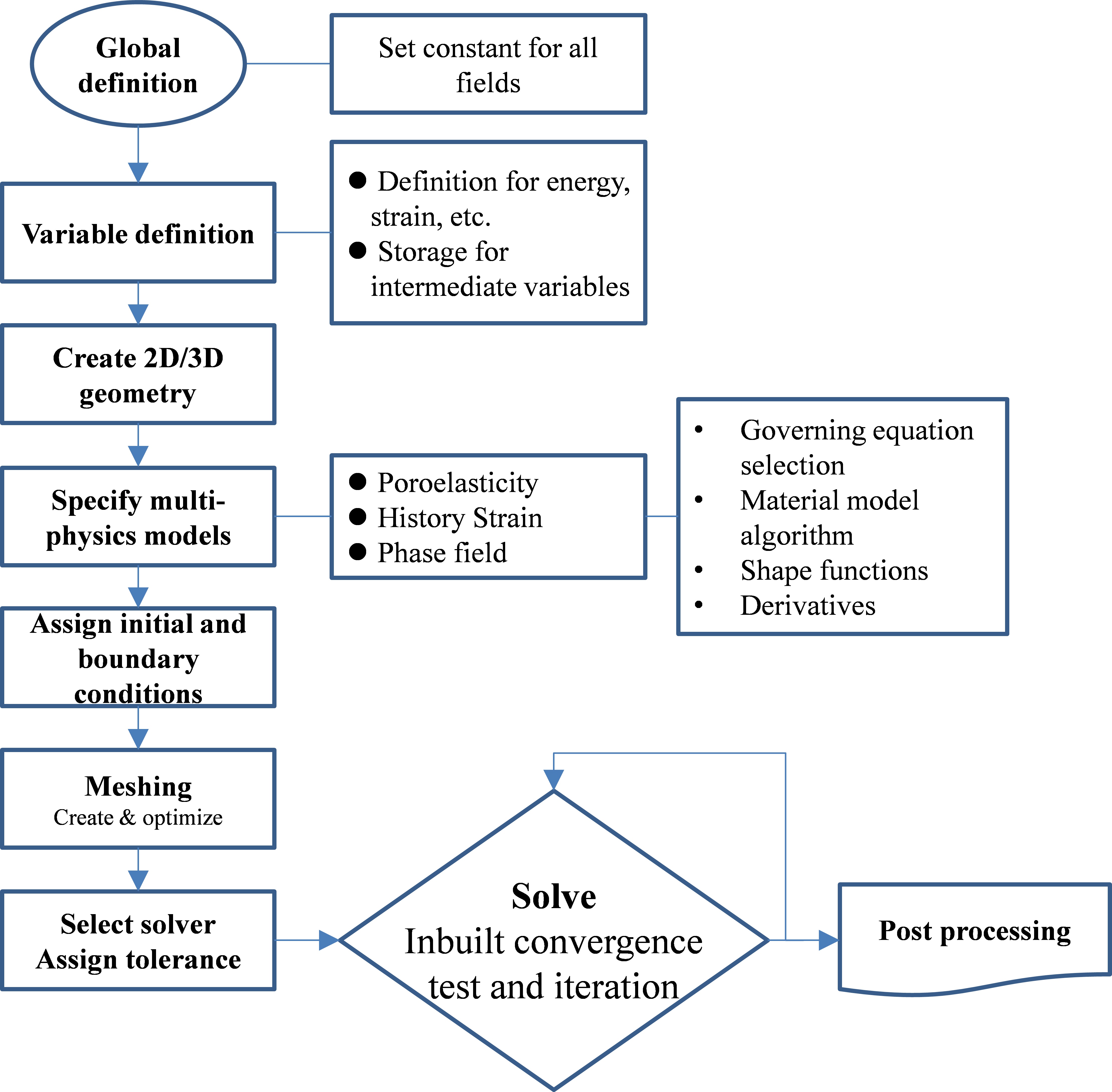}
	\caption{Comsol implementation of phase field modeling for fluid-driven crack problems}
	\label{Comsol implementation of phase field modeling for fluid-driven crack problems}
	\end{figure}

Since the three fields are highly coupled, the iteration is difficult to converge when the porous medium starts to fracture. The iteration steps will increase to a huge number. In Comsol, the Anderson acceleration method is used for nonlinear convergence acceleration by the information from previous Newton iterations \citep{comsol2005comsol}. Therefore, we use the Anderson acceleration method to accelerate convergence. In this paper, the dimension of iteration space field is set as more than 50 to control the number of iteration increments.  The maximum number of iteration in one time step is set as 50 in our simulations. Finally, our implementation procedure of the phase field modeling for the fluid-driven crack propagation in Comsol is shown in Fig. \ref{Comsol implementation of phase field modeling for fluid-driven crack problems}.

\section {Verification of the proposed approach}
\subsection{A single-phasic solid subjected to tension}

\begin{figure}[htbp]
	\centering
	\includegraphics[width = 8cm]{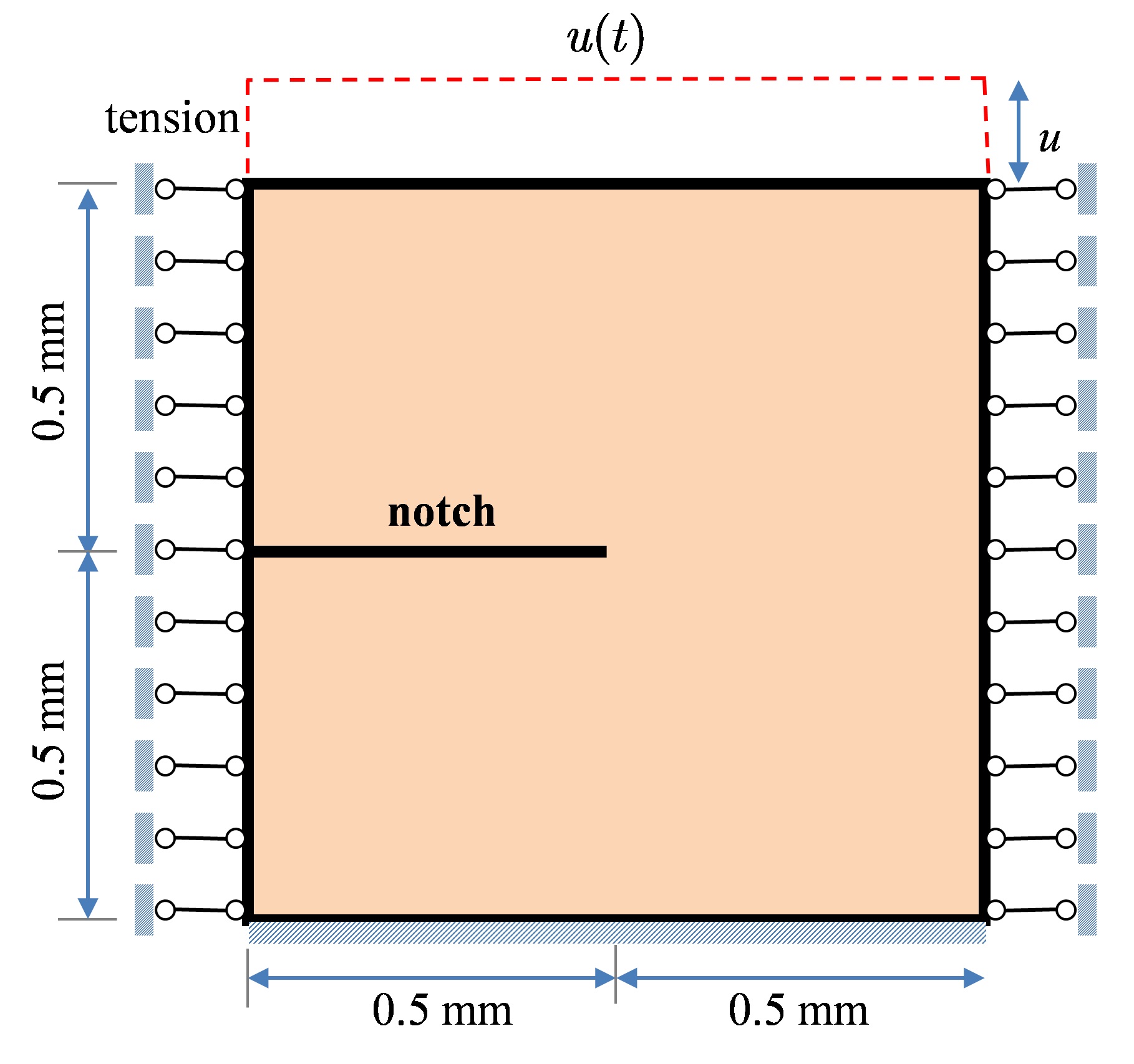}
	\caption{Geometry and boundary conditions of the single-phasic solid subjected to tension}
	\label{Geometry and boundary conditions of the single-phasic solid subjected to tension}
	\end{figure}

The first example is a single-phasic solid subjected to tension. This example has been tested by \citet{miehe2010phase, miehe2010thermodynamically}, \citet{ liu2016abaqus}, \citet{hesch2014thermodynamically}, \citet{ mikelic2013phase}, and \citet{ehlers2017phase}. The geometry and boundary conditions in Fig. \ref{Geometry and boundary conditions of the single-phasic solid subjected to tension} are used and the calculation parameters are listed in Table \ref{Parameters for the single-phasic solid subjected to tension}. These parameters are taken from \citep{ liu2016abaqus}.

	\begin{table}[htbp]
	\caption{Parameters for the single-phasic solid subjected to tension}
	\label{Parameters for the single-phasic solid subjected to tension}
	\centering
	\begin{tabular}{p{10pt}p{80pt}p{10pt}p{200pt}p{10pt}p{100pt}}
	\hline
	$\mu$ & 80.77 GPa & $\lambda$ &121.15 GPa & $G_c$ &2700 N/m\\
	$k$ & $1\times10^{-9}$ & $l_0$ &$7.5\times10^{-3}$ mm,  $1.5\times10^{-2}$ mm&   & \\
	\hline
	\end{tabular}
	\end{table}

\begin{figure}[htbp]
	\centering
	\subfigure[$l_0=1.5\times10^{-2}$ mm]{\includegraphics[width = 5cm]{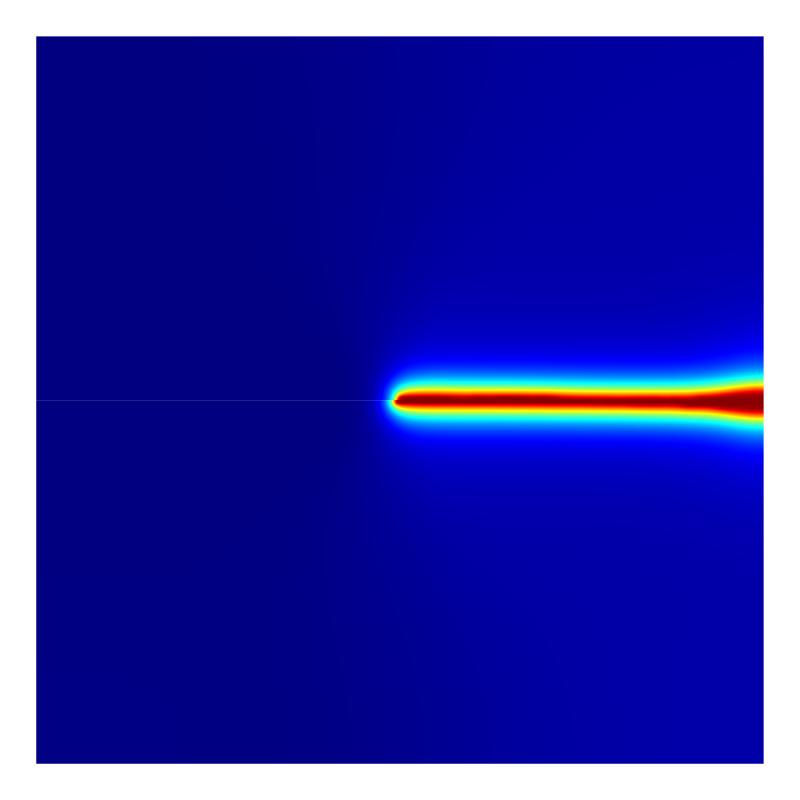}}
	\subfigure[$l_0=7.5\times10^{-3}$ mm]{\includegraphics[width = 5cm]{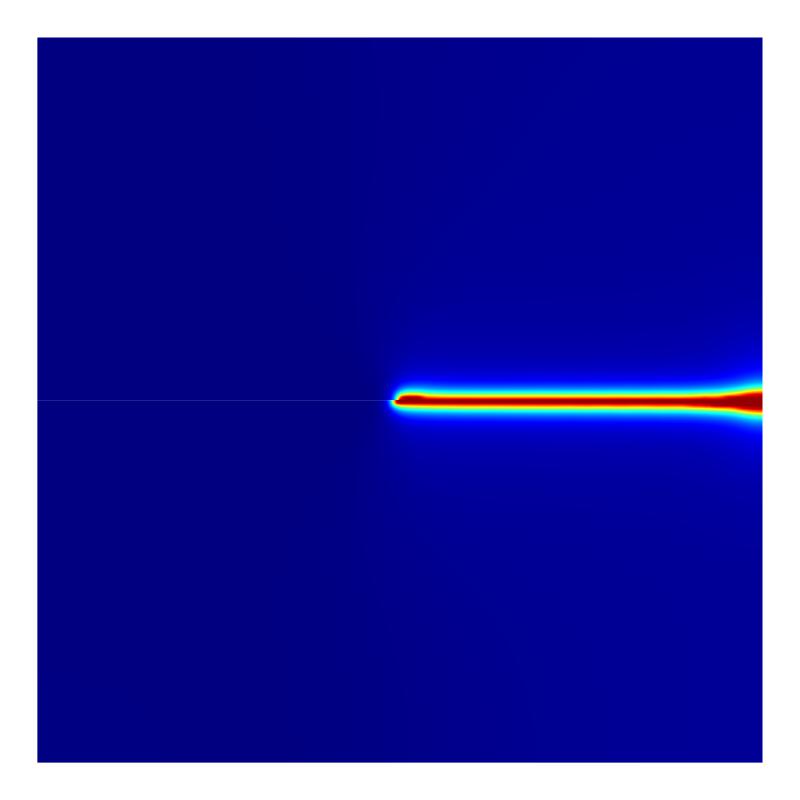}}
	\caption{Crack pattern of the single-phasic solid subjected to tension}
	\label{Crack pattern of the single-phasic solid subjected to tension}
	\end{figure}

\begin{figure}[htbp]
	\centering
	\includegraphics[width = 8cm]{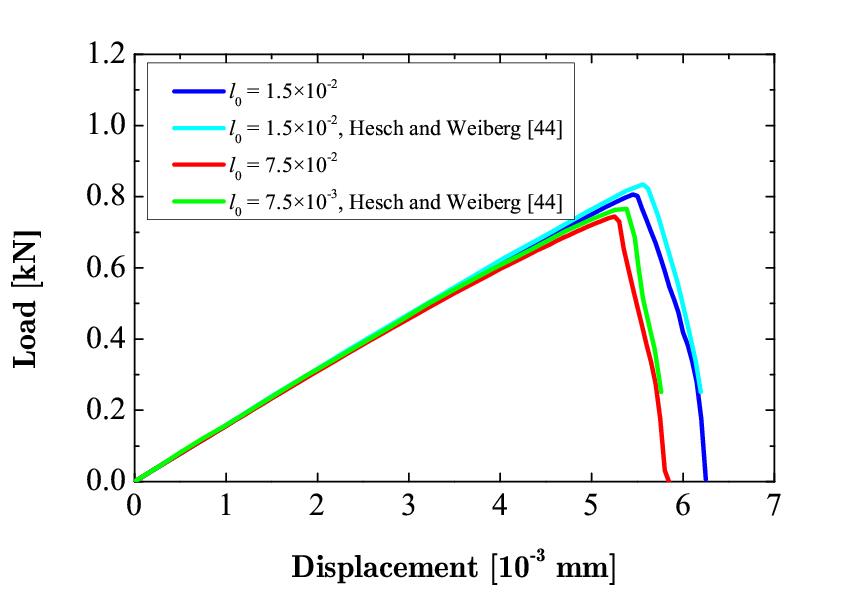}
	\caption{Load-displacement curve of the single-phasic solid subjected to tension}
	\label{Load-displacement curve of the single-phasic solid subjected to tension}
	\end{figure}

Note that the gravity in all the examples  is neglected. We test the notched plate without a pressure ($p=0$). Thus, calculation for the pressure is removed from the segregated steps. The vertical displacement $u(t)=vt$ is prescribed on the top edge of the plate with $v=1\times10^{-6}$ mm/s. We use 64516 uniform Q4 elements to discretize the solid with the element size $h$ around $3.96\times 10^ {-3}$ mm. We use the same element size as in \citet{hesch2014thermodynamically} so that the results are directly comparable. We apply a time step $\Delta t =10$ s for the first 4500 s, afterwards, $\Delta t =1$ s for the remaining time.  This example has been solved within 4 h 20 min  for $l_0 = 1.5 \times 10^{-2}$ mm  and 4 h 11 min  for $l_0 = 7.5 \times 10^{-3}$ mm  on two I5-6200U CPUs. As shown in Fig. \ref{Crack pattern of the single-phasic solid subjected to tension}, a horizontal crack propagates through the solid as described in \citep{miehe2010phase, miehe2010thermodynamically,liu2016abaqus,hesch2014thermodynamically,mikelic2013phase,ehlers2017phase}. In addition, Fig. \ref{Load-displacement curve of the single-phasic solid subjected to tension} gives the load-displacement curve on the top edge of the solid. The calculated curves are in good agreement with the results by \citet{hesch2014thermodynamically}. Only marginal differences are observed because different algorithms are used.

\subsection{A 2D specimen subjected to an increasing internal pressure}

\begin{figure}[htbp]
	\centering
	\subfigure[The whole specimen]{\includegraphics[height = 6cm]{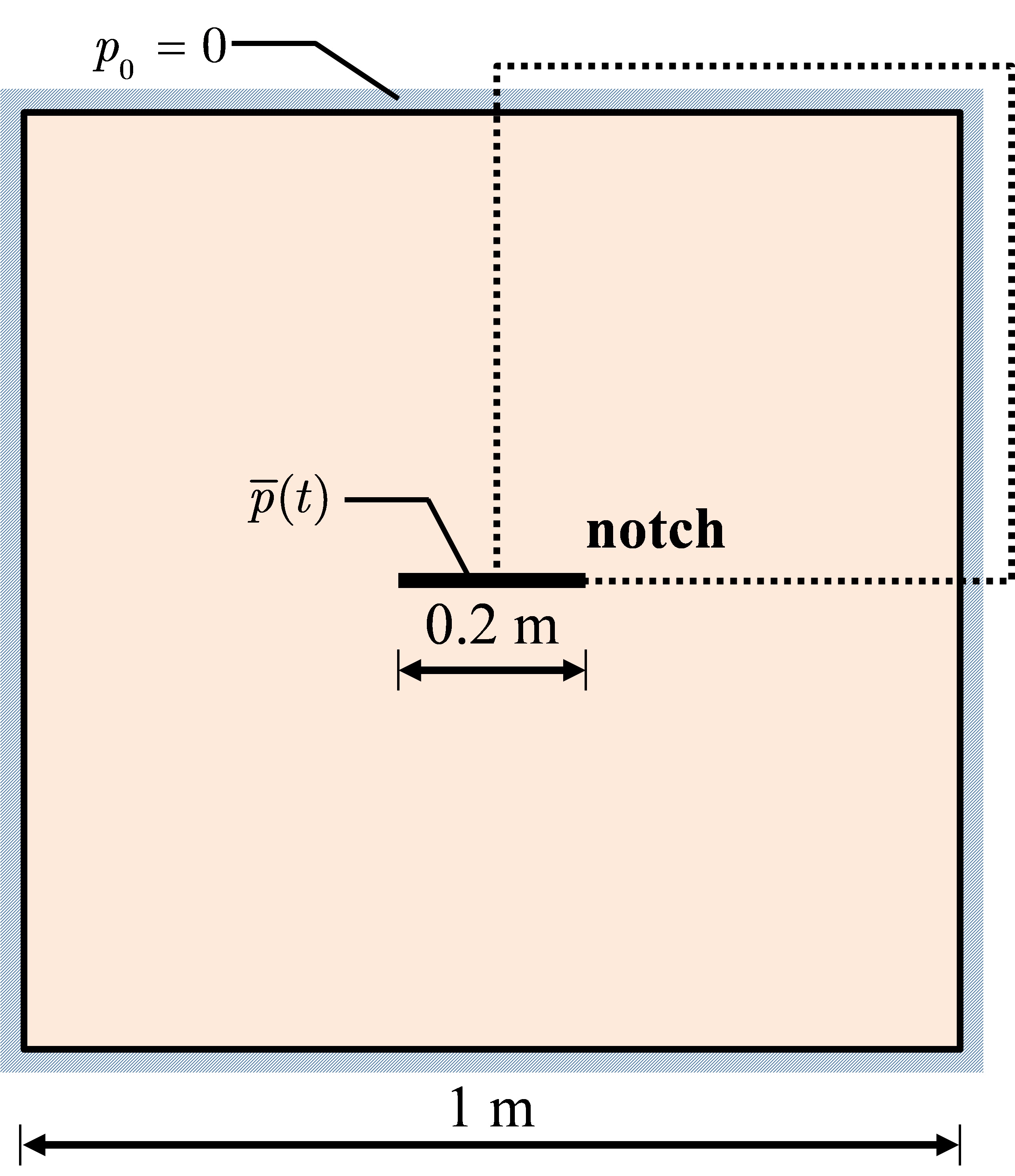}}
	\subfigure[Top-right quarter of the specimen]{\includegraphics[height = 6cm]{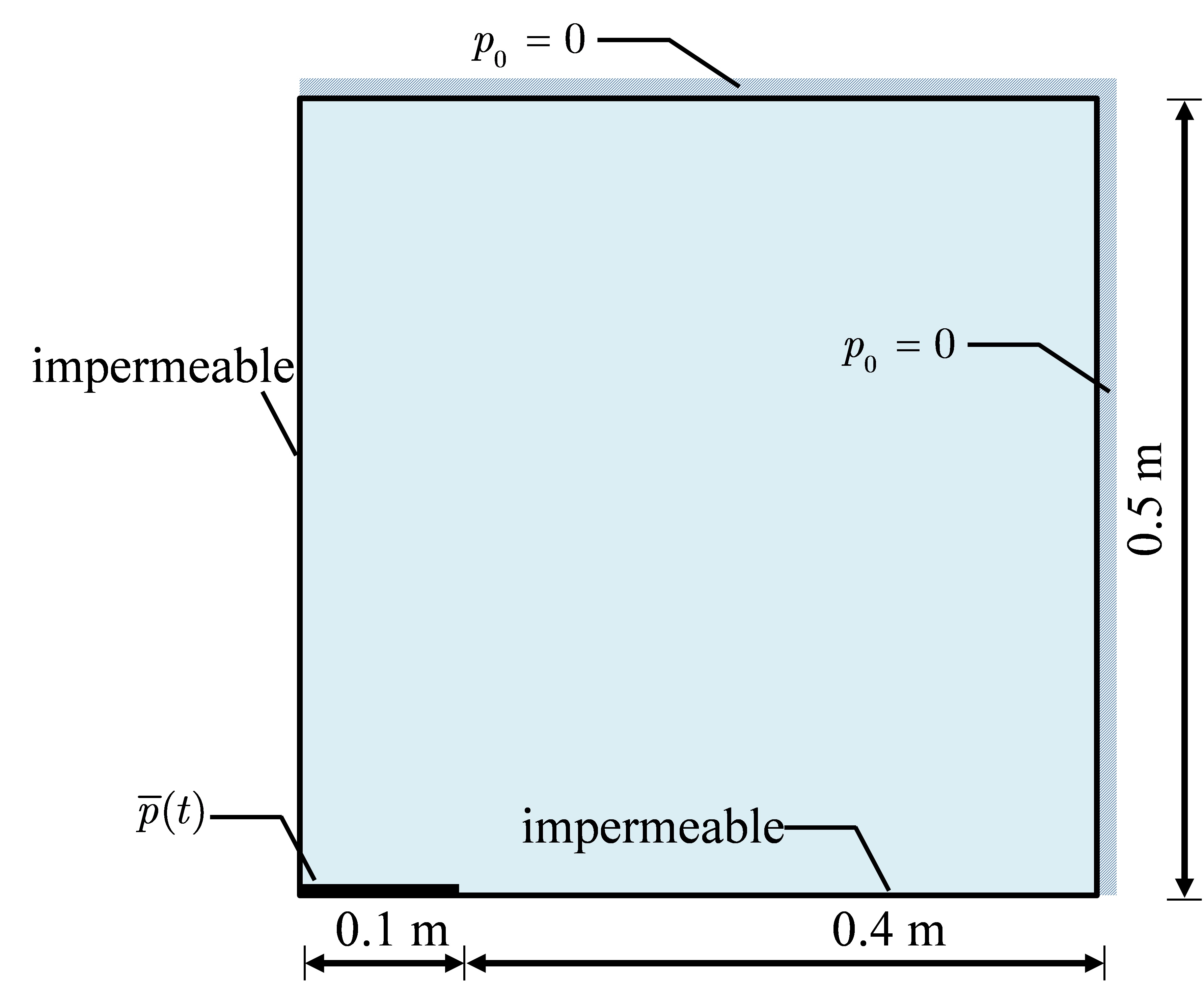}}
	\caption{Geometry and boundary conditions of the notched specimen subjected to internal pressure}
	\label{Geometry and boundary conditions of the notched specimen subjected to internal pressure}
	\end{figure}

The second example simulates the fluid-driven crack propagation of a saturated notched specimen subjected to an increasing internal pressure along the notch. The geometry and boundary conditions are depicted in Fig. \ref{Geometry and boundary conditions of the notched specimen subjected to internal pressure}. We use this example to verify the feasibility of the phase filed model for fracture  in poroelastic media. Due to the double symmetry in Fig. \ref{Geometry and boundary conditions of the notched specimen subjected to internal pressure}a, only the top-right quarter of the specimen is modeled, cf. Fig. \ref{Geometry and boundary conditions of the notched specimen subjected to internal pressure}b. AIl the parameters are listed in Table \ref{Parameters for the notched specimen subjected to internal pressure}.

	\begin{table}[htbp]
	\small
	\caption{Parameters for the notched specimen subjected to internal pressure}
	\label{Parameters for the notched specimen subjected to internal pressure}
	\centering
	\begin{tabular}{llllllllll}
	\hline
	$\mu$ & 80.77 GPa & $\lambda$ &121.15 GPa & $G_c$ &2700 N/m & $k$ &$1\times10^{-9}$ & $l_0$ & $4\times10^{-3}$ m\\
	$c_1$ & 0.4 & $c_2$ & 1.0 & $\varepsilon_{pr}$ & 0.002 & $\rho_{r}$,  $\rho_{f}$& $1.0\times10^{3}$ kg/m$^3$  & $\alpha_r$ &0.002 \\ $q_r$ & 0 & $q_f$ & 0 & $K_r$ & $1\times10^{-15}$ m$^2$ & $K_f$ & $1.333\times10^{-6}$ m$^2$& &\\
$c_r$ & $1\times10^{-8}$ 1/Pa & $c_f$ & $1\times10^{-8}$ 1/Pa & $\mu_r$ & $1\times10^{-3}$ Pa$\cdot$s & $\mu_f$ & $1\times10^{-3}$ Pa$\cdot$s& &\\
	\hline
	\normalsize
	\end{tabular}
	\end{table}

We discretize the domain with uniformly spaced Q4 elements. Two mesh levels are used: Mesh 1 with the element size $h=2\times10^{-3}$ m and Mesh 2 with $h=1\times10^{-3}$ m. A total of 2000 time steps are used with the time increment $\Delta t=1$ s. For Meshes 1 and 2, the time cost on two I5-6200 CPUs is 19 h 18 min and 53 h 23 min, respectively. Analytical solutions are used to verify the numerical results. Considering a single pre-existing crack with a length of $2l_c$ in $y=0$ plane, the displacement driven by the pressure in the $y$ direction under the plain strain assumption is given by \citep{sneddon1969crack} as 
	\begin{equation}
	u^+(0,x)=\frac{2pl_c}{E_p}\left(1-\frac{x^2}{l_c^2}\right)^{1/2}
	\label{u+}
	\end{equation}

\noindent where $E_p=E/(1-\nu^2)$ is the plane strain Young's modulus and $E$ and $\nu$ are the Young's modulus and Poisson's ratio, respectively.

According to \citet{yoshioka2016variational} and \citet{mikelic2013phase}, the fracture volume is $V=2\pi pl_c/E_p$. Thus, the dissipated energy $E_b$ are calculated by
	\begin{equation}
	E_b = \frac{1}{2}pV=\frac{\pi l_c^2p^2}{E_p}
	\end{equation}

The energy release rate $G_c$ are calculated as $E_b/(2l_c)$ according to Griffith's theory. Therefore, the critical pressure for fracture propagation is
	\begin{equation}
	p_c= \left(\frac{2E_pG_c}{\pi l_0}\right)^{\frac{1}{2}}
	\label{pc}
	\end{equation}

\begin{figure}[htbp]
	\centering
	\subfigure[Mesh 1]{\includegraphics[width = 10cm]{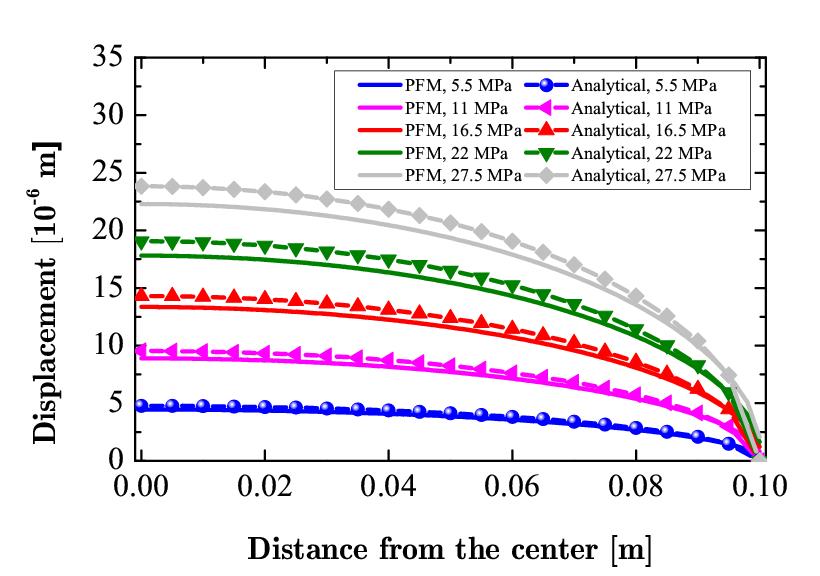}}\\
	\subfigure[Mesh 2]{\includegraphics[width = 10cm]{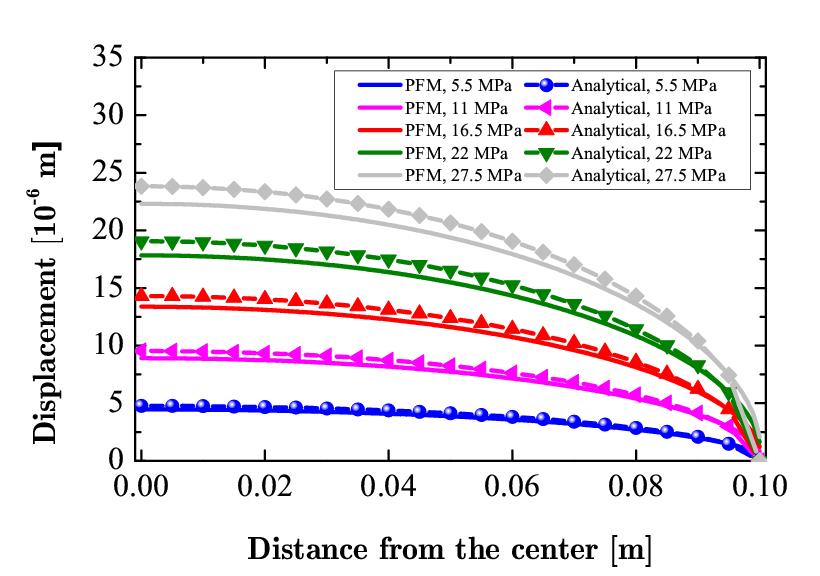}}
	\caption{Comparison of the displacement along the notch between the present results and the analytical solution}
	\label{Comparison of the displacement along the notch between the present results and the analytical solution}
	\end{figure}

\begin{figure}[htbp]
	\centering
	\includegraphics[width = 10cm]{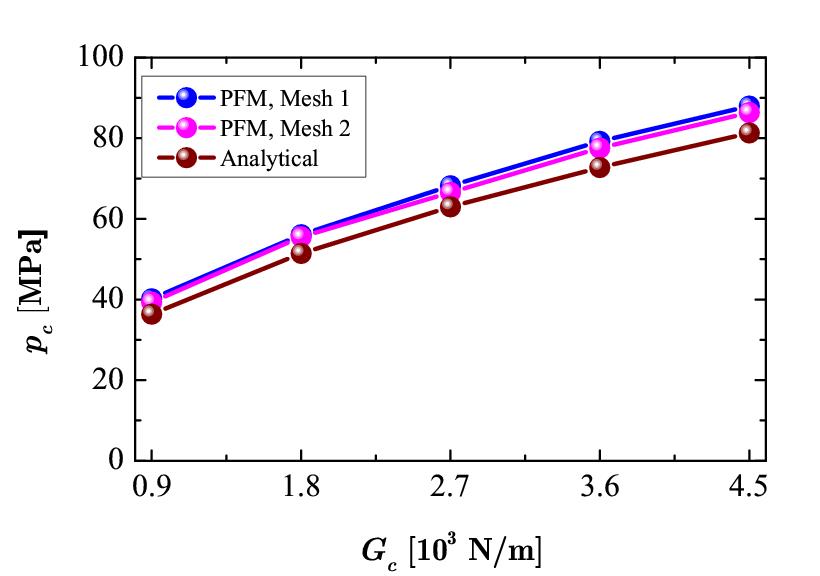}
	\caption{Comparison of the critical pressure for fracture propagation between the present results and the analytical solution}
	\label{Comparison of the critical pressure for fracture propagation between the present results and the analytical solution}
	\end{figure}

Figure \ref{Comparison of the displacement along the notch between the present results and the analytical solution} compares the displacement along the notch between the present results and the analytical solution. Both the numerical results and analytical solution are in good agreement. Moreover, the displacements given by the analytical solution are larger than those by the phase field model. The critical pressure for fracture initiation between the present results and analytical solution is shown in Fig. \ref{Comparison of the critical pressure for fracture propagation between the present results and the analytical solution}. The critical pressure increases with increasing $G_c$ and the numerical results are larger than those obtained by the analytical solution. In addition, a finer mesh (Mesh 2) will lead to a closer critical pressure to the analytical solution. The slight difference between the simulated critical pressure and the analytical solution occurs because the analytical solution is derived based on a sharp crack shape assumption with fully impermeable crack surfaces. On the other hand, the pre-existing notch in the PFM is a smooth representation of the newly formed crack around the notch tip and the notch surface is slightly permeable (with very low permeability and porosity). 

\begin{figure}[htbp]
	\centering
		\begin{tabular}{lllcll}
		\centering
		$\textbf{a}$ & $t=1250$ s & $\textbf{b}$ & $t=1550$ s &$\textbf{c}$ & $t=1750$ s\\
		&\includegraphics[width = 5cm]{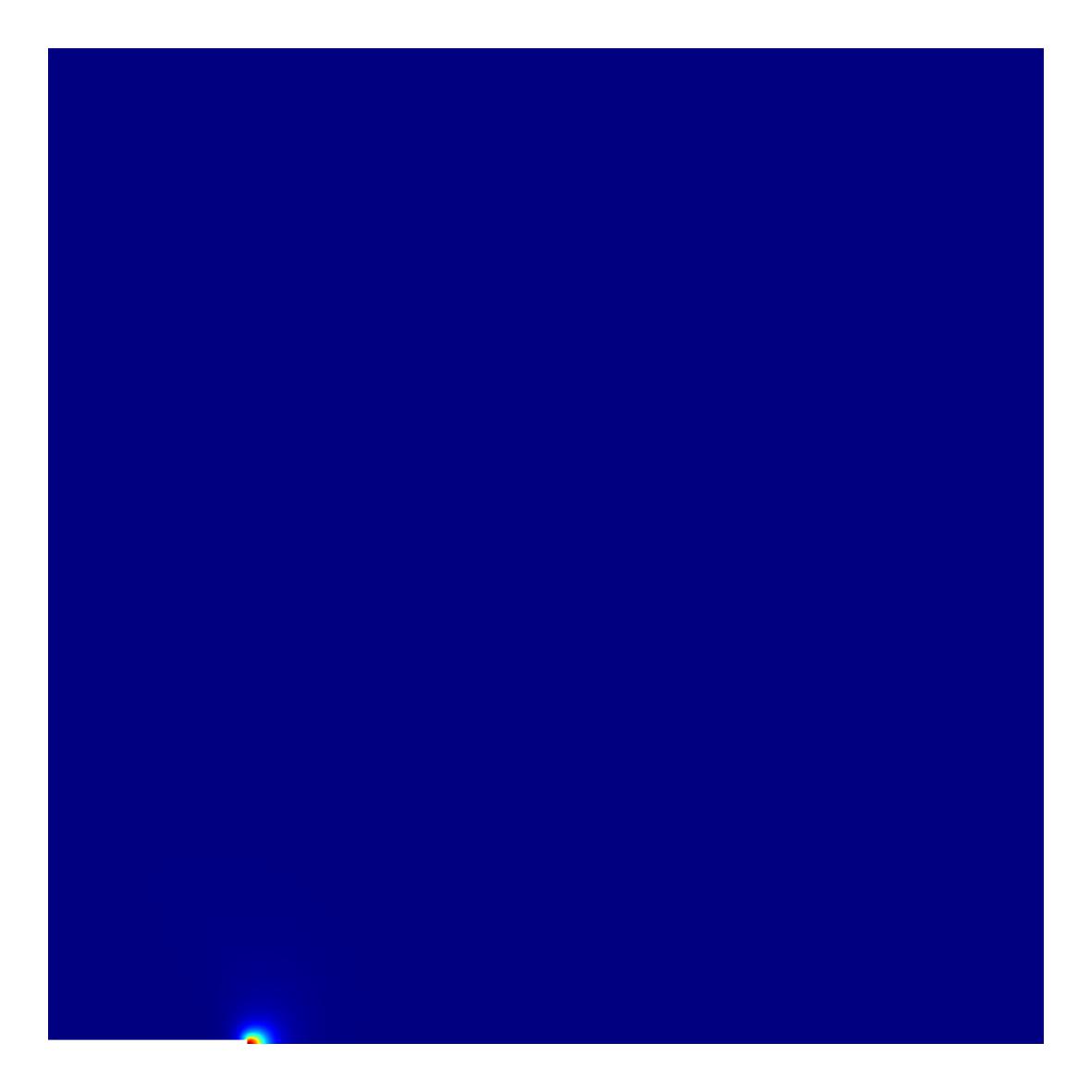}&&\includegraphics[width = 5cm]{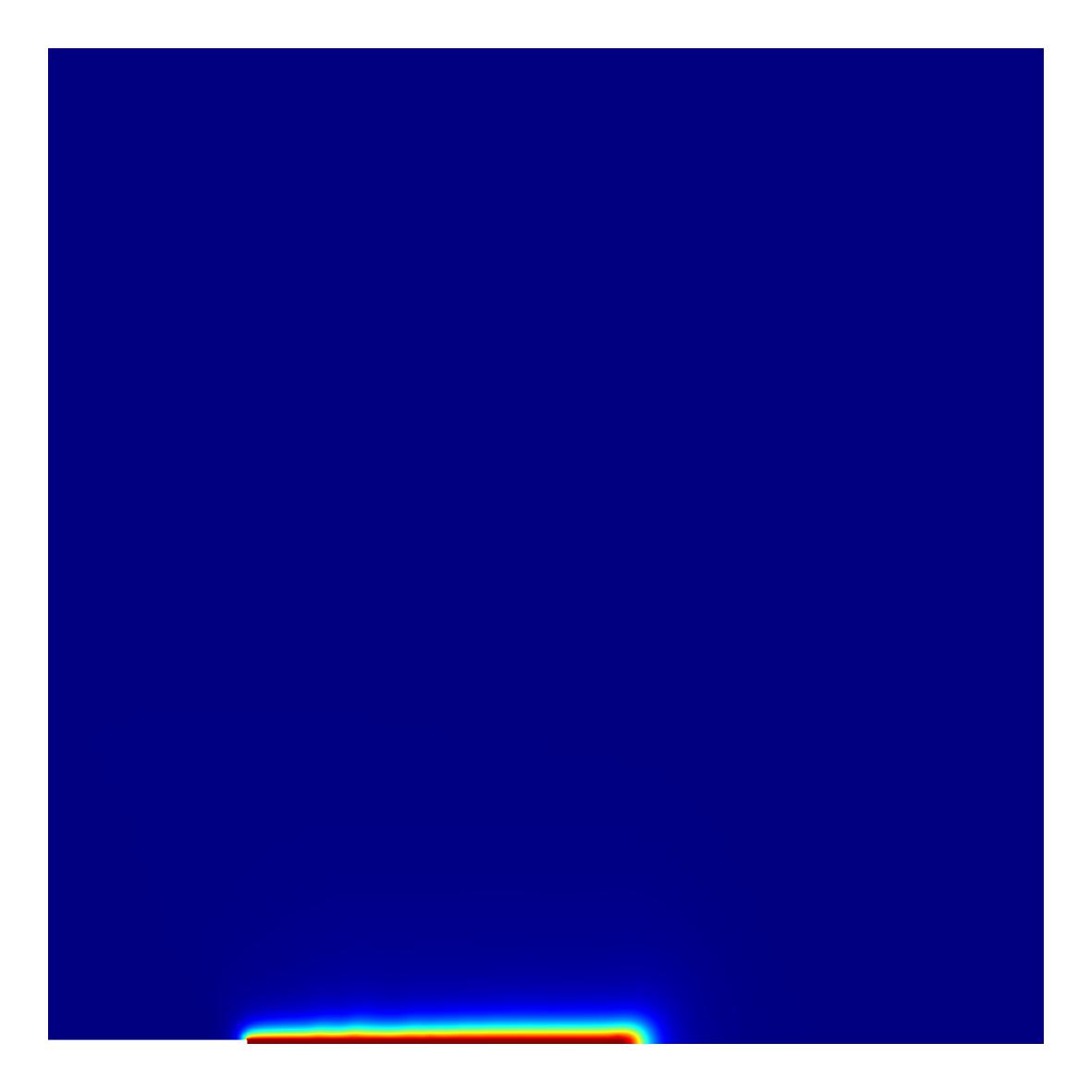}&&\includegraphics[width = 5cm]{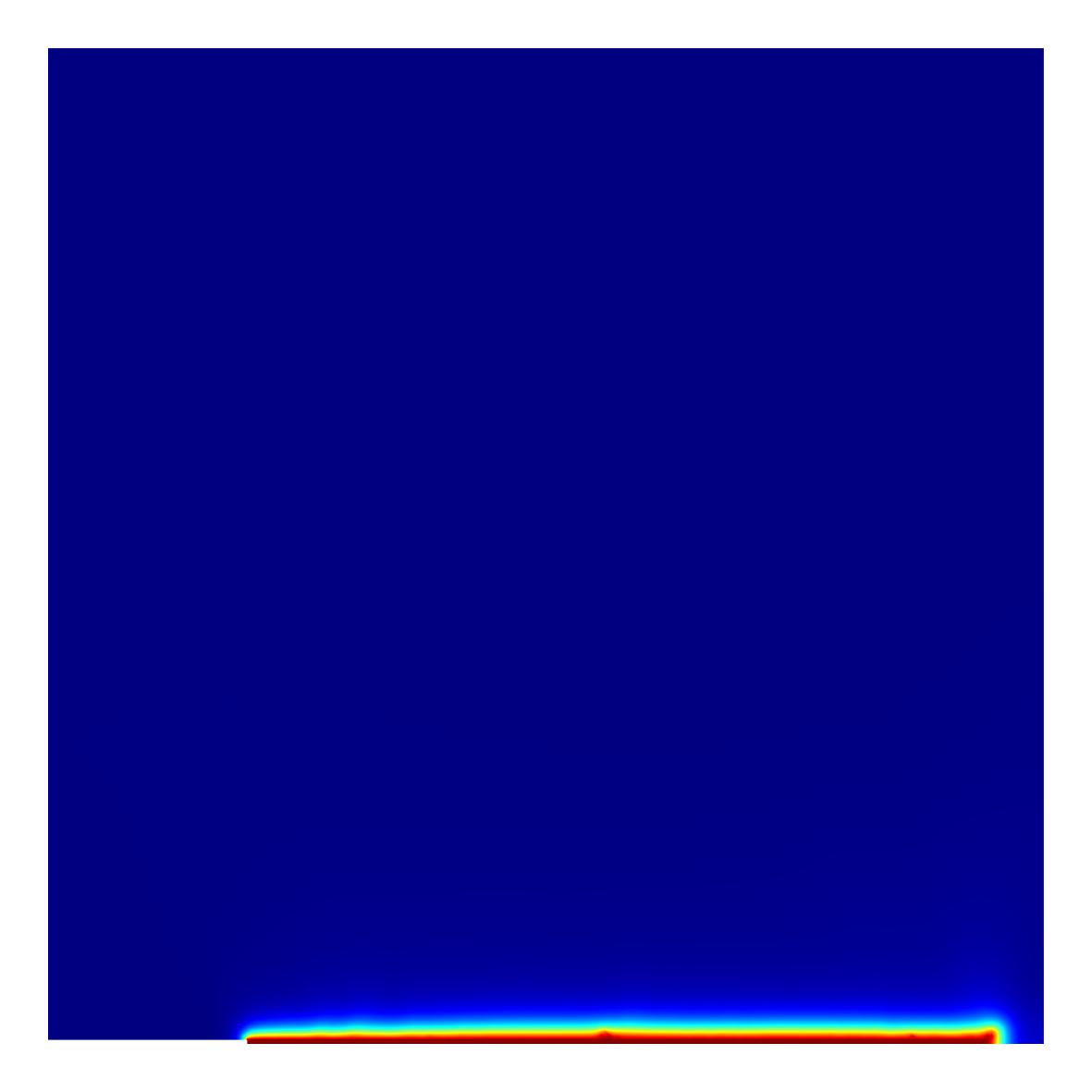}\\&&$\phi$&\includegraphics[width = 5cm]{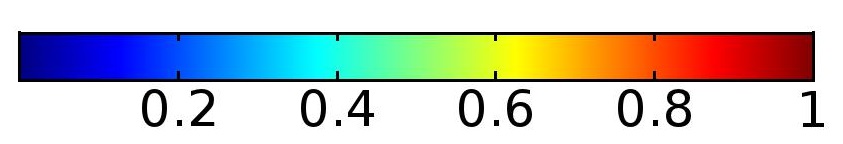}&\\
$\textbf{d}$ & & $\textbf{e}$ & &$\textbf{f}$ &\\
		&\includegraphics[width = 5cm]{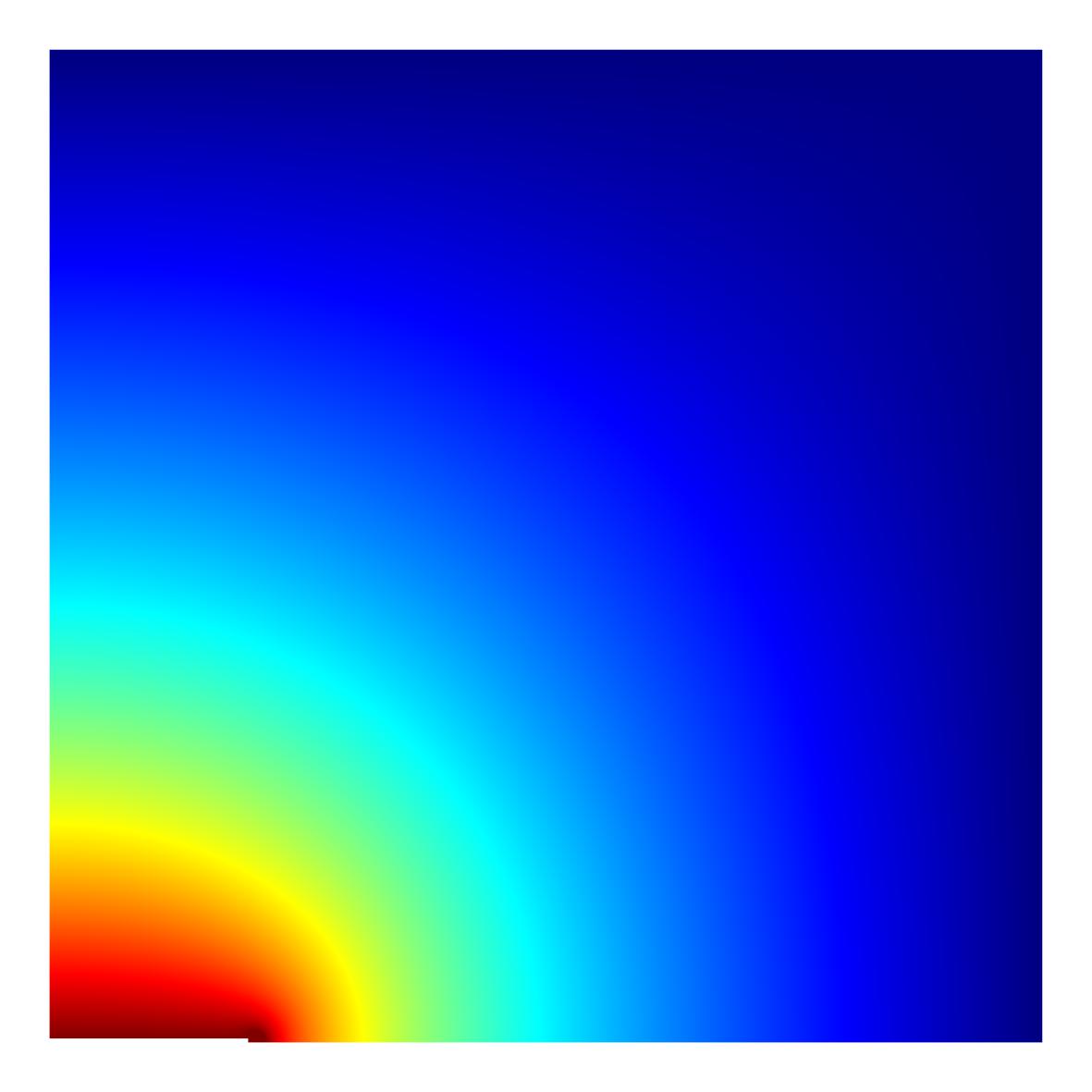}&&\includegraphics[width = 5cm]{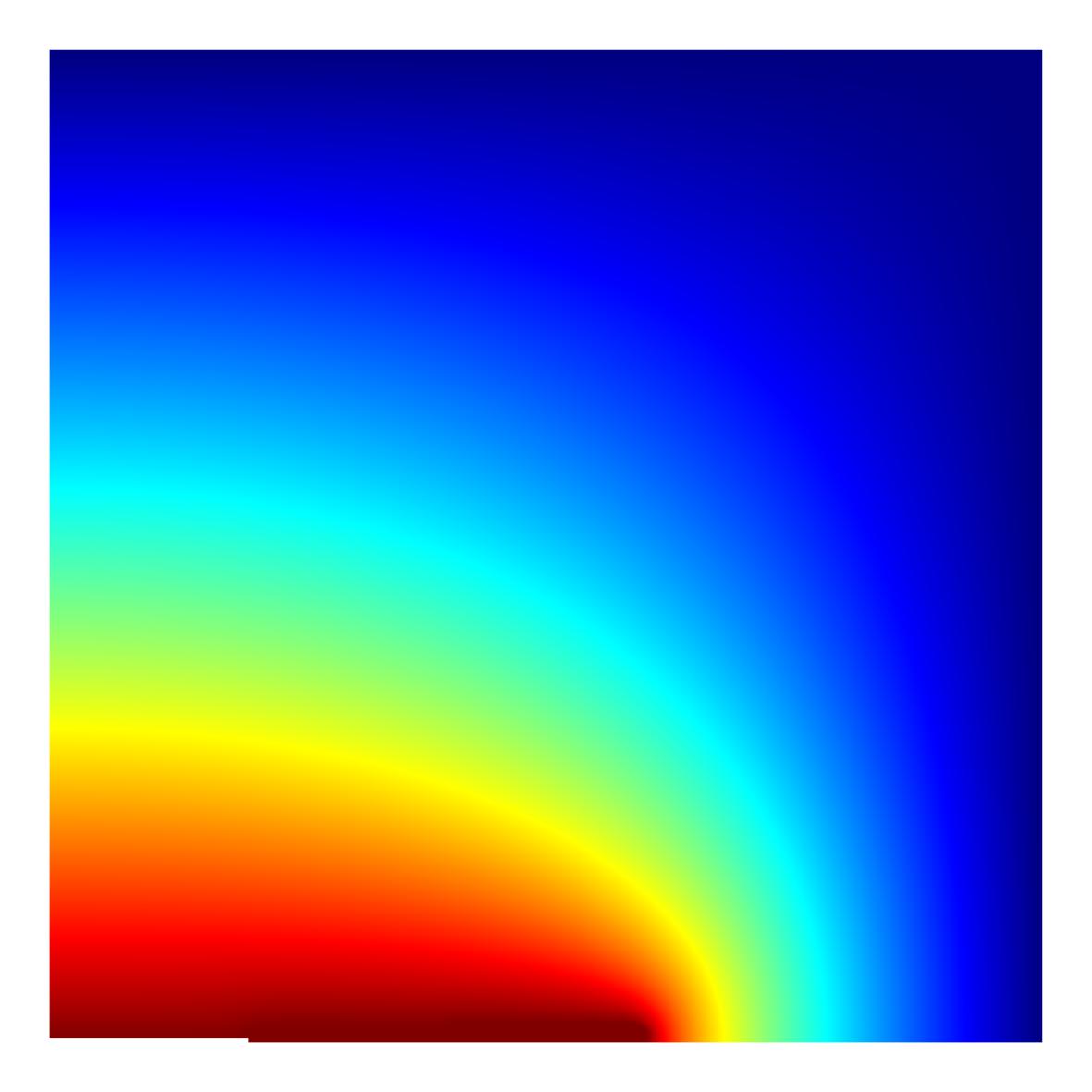}&&\includegraphics[width = 5cm]{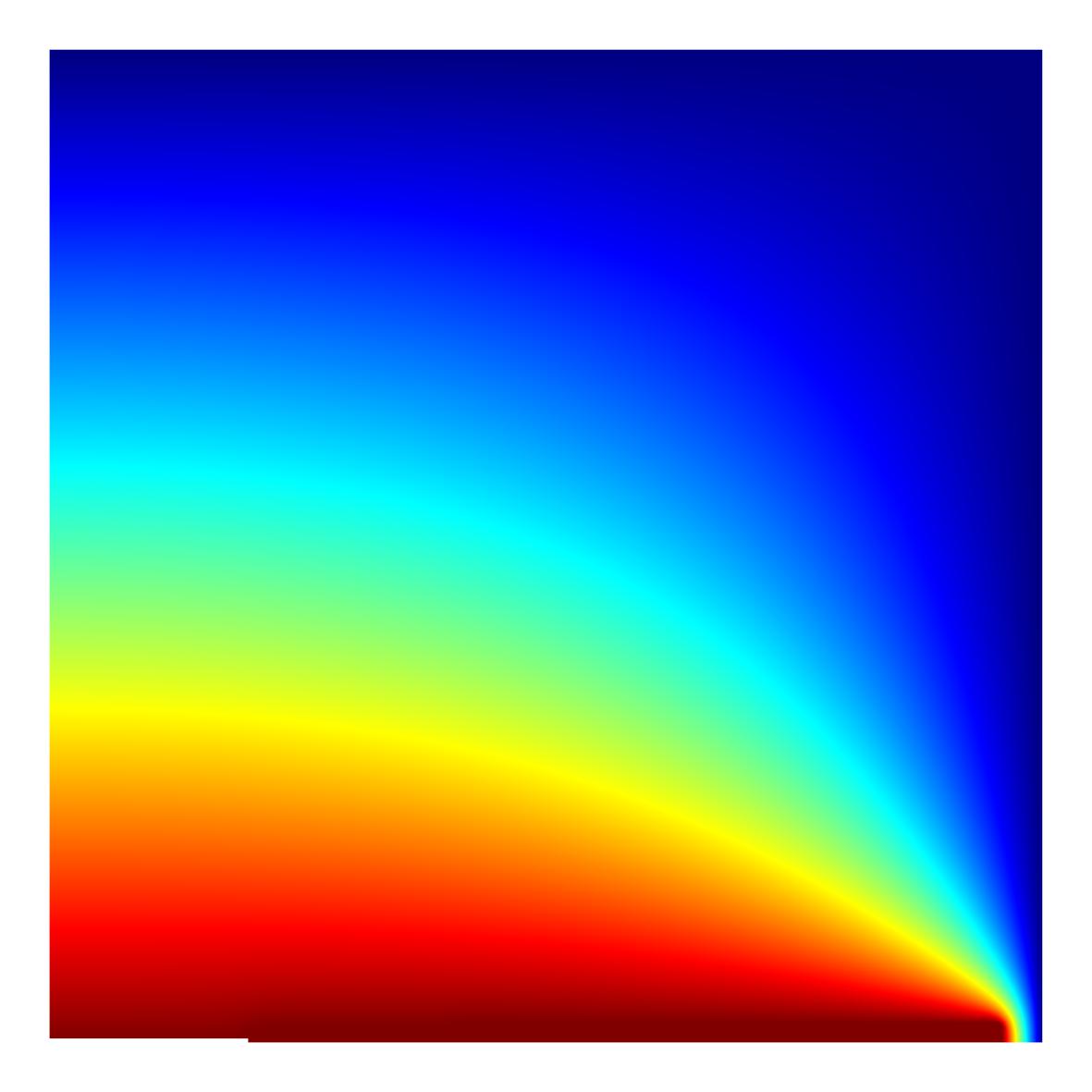}\\&&$p/\bar{p}$&\includegraphics[width = 5cm]{Figure12b2.jpg}&\\
		\end{tabular}
	\caption{Phase field (a)-(c) and fluid pressure (d)-(f) of the specimen subjected to an increasing internal pressure for Mesh 1}
	\label{Phase field (a)-(c) and fluid pressure (d)-(f) of the specimen subjected to an increasing internal pressure}
	\end{figure}

With the increase in the applied pressure, the width of the notch starts to increase and the fracture propagates along the horizontal direction. Figure \ref{Phase field (a)-(c) and fluid pressure (d)-(f) of the specimen subjected to an increasing internal pressure} shows the evolution of the phase field $\phi$ in (a)-(c) and the fluid pressure in (d)-(f). No significant variation in the width of the crack is observed as the time $t$ goes. In addition, the maximum pressure occurs inside the crack although radial distribution of the fluid pressure along the crack is also observed. The pressure gradient is much smaller along the crack than perpendicular to the crack. Therefore, as observed, the crack almost has a uniformly distributed fluid pressure. 

\begin{figure}[htbp]
	\centering
	\includegraphics[height = 6cm]{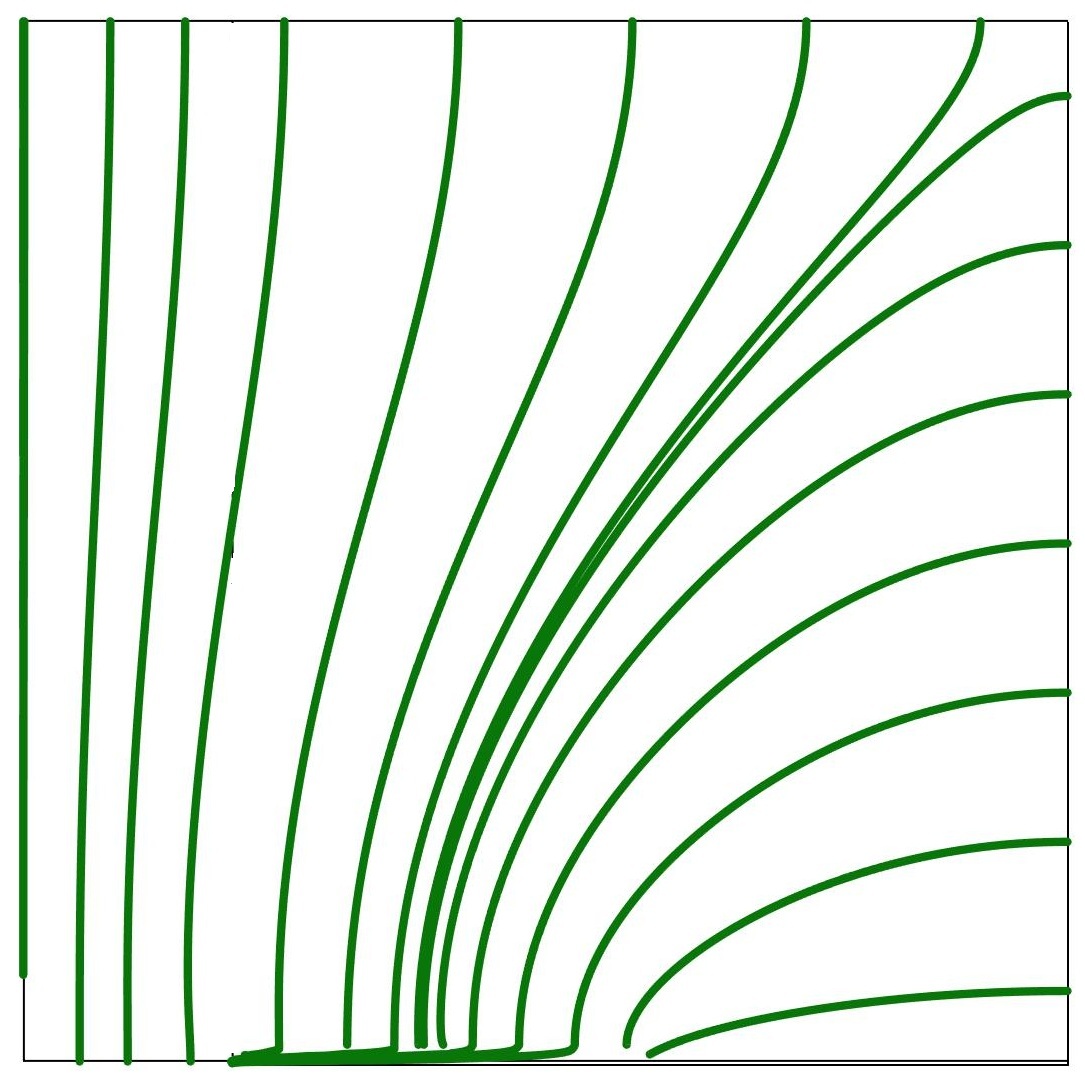}
	\caption{Streamlines of the fluid flow at $t = 1550$ s for Mesh 1}
	\label{Streamlines of the fluid flow at t = 1550 s}
	\end{figure}

Figure \ref{Streamlines of the fluid flow at t = 1550 s} gives the streamlines of the fluid flow at $t=1550$ s for Mesh 1. Two directions of fluid flow are observed. The major one is the flow along the propagating crack. The other flow is initially perpendicular to the notch and the crack and finally perpendicular to the boundaries.

\section{2D and 3D examples of internal fluid injection}
In this section, 2D and 3D examples of specimens subjected to internal fluid injection are presented to demonstrate  capability of the proposed modeling approach. Pre-existing cracks are modeled by introducing an initial history field and the source term $q_{F}=$ 100 kg/(m$^3\cdot \textrm s$) is used for all examples. The crack propagation is therefore driven by the fluid injection in the pre-existing cracks. The parameters for all calculations can be found in Table \ref{Parameters for the notched specimen subjected to internal pressure}.

\subsection{Two perpendicular propagating cracks in 2D}

\begin{figure}[htbp]
	\centering
	\includegraphics[width = 8cm]{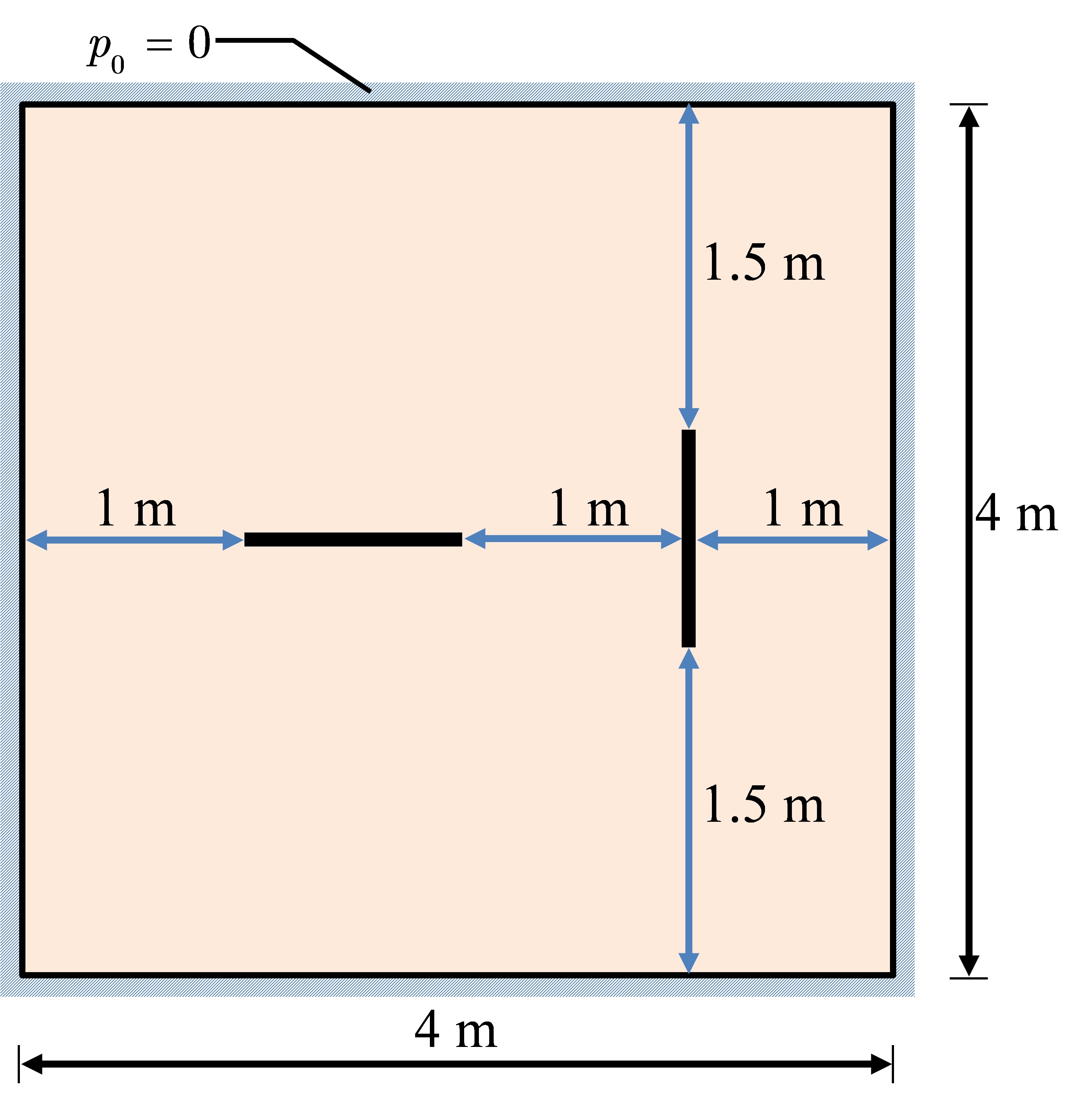}
	\caption{Geometry and boundary condition of the two perpendicular pre-existing cracks}
	\label{Geometry and boundary condition of the two perpendicular pre-existing cracks}
	\end{figure}

Let us consider a crack propagation example of two perpendicular pre-existing cracks. Fluid is injected into both cracks and the geometry and boundary conditions are shown in Fig. \ref{Geometry and boundary condition of the two perpendicular pre-existing cracks}. The length of the cracks is 1 m. The element size is $h=$ $2\times 10^{-2}$ m. A total of 1160 time steps are used with the time increment $\Delta t=0.01$ s. The time cost on two I5-6200 CPUs is 12 h 49 min for this example.

\begin{figure}[htbp]
	\centering
	\subfigure[$t=0$]{\includegraphics[height = 5cm]{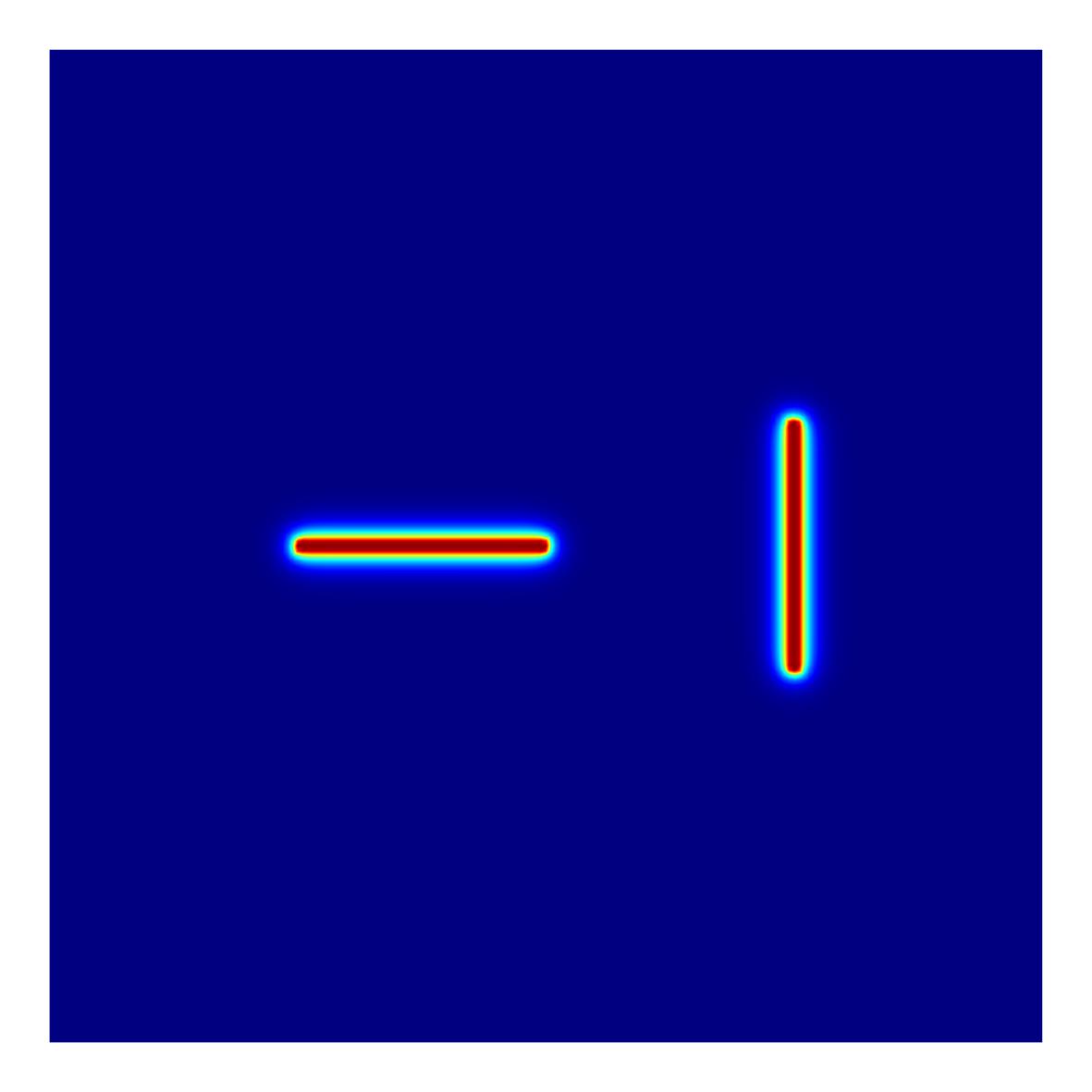}}
	\subfigure[$t=6$ s]{\includegraphics[height = 5cm]{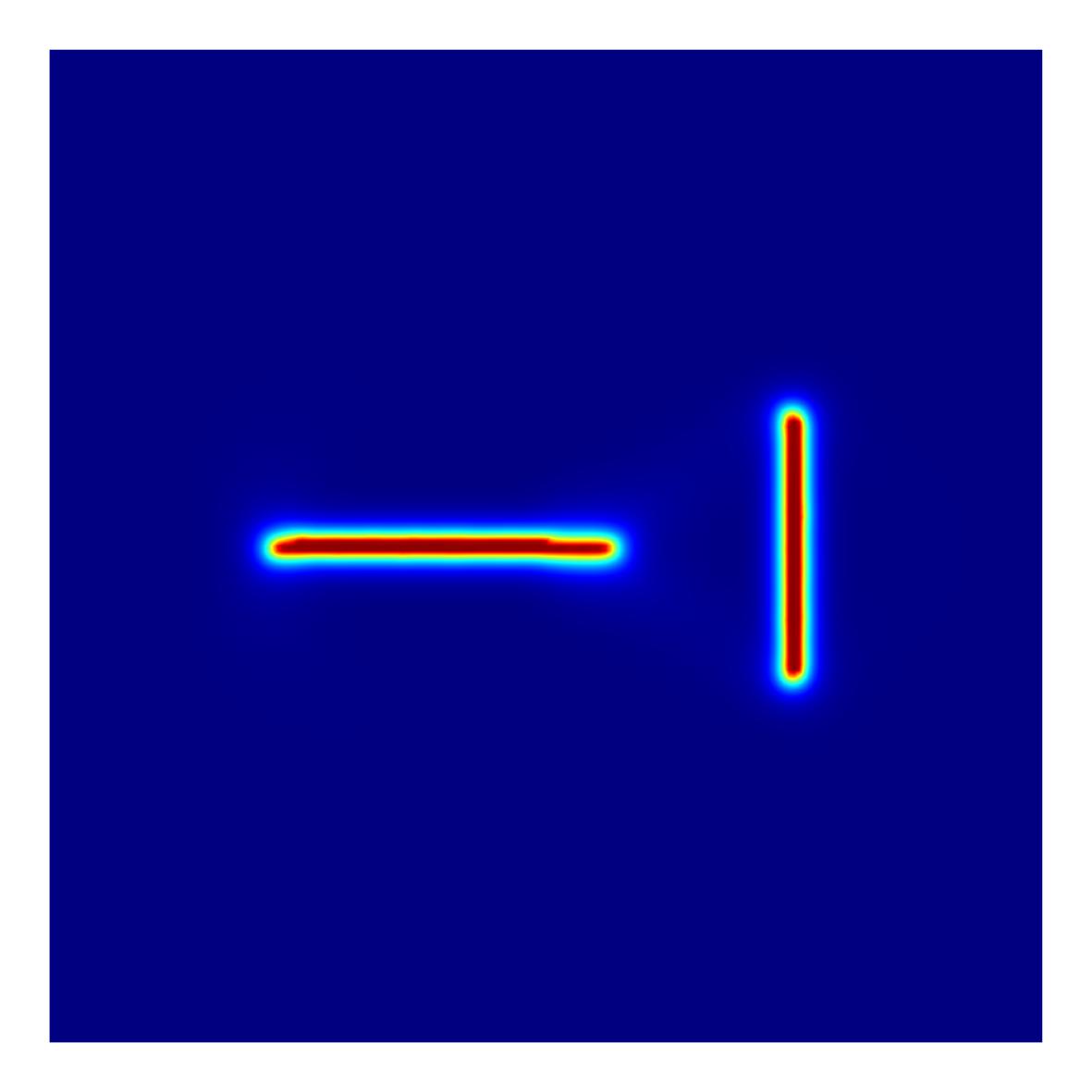}}
	\subfigure[$t=6.8$ s]{\includegraphics[height = 5cm]{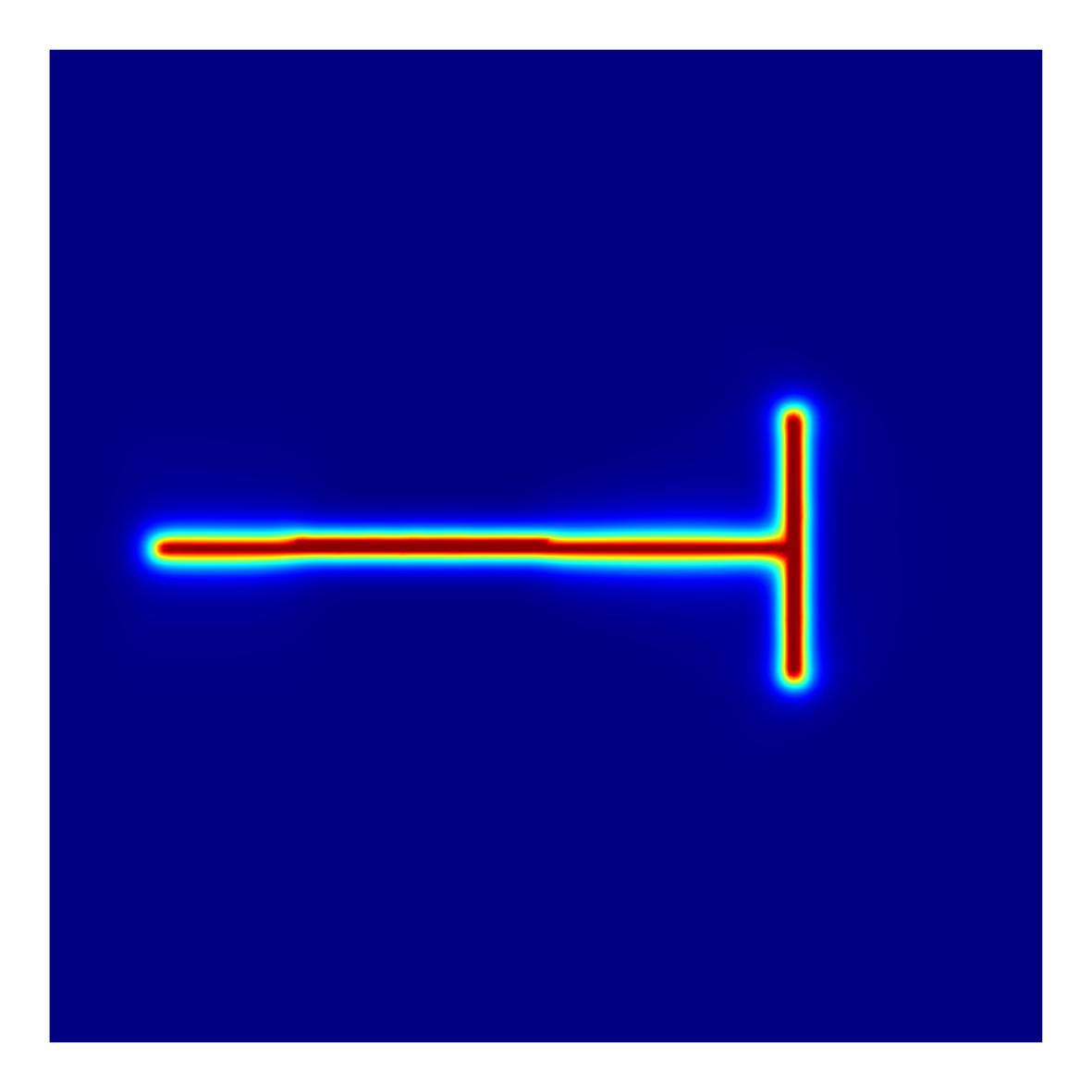}}\\
	\subfigure[$t=8$ s]{\includegraphics[height = 5cm]{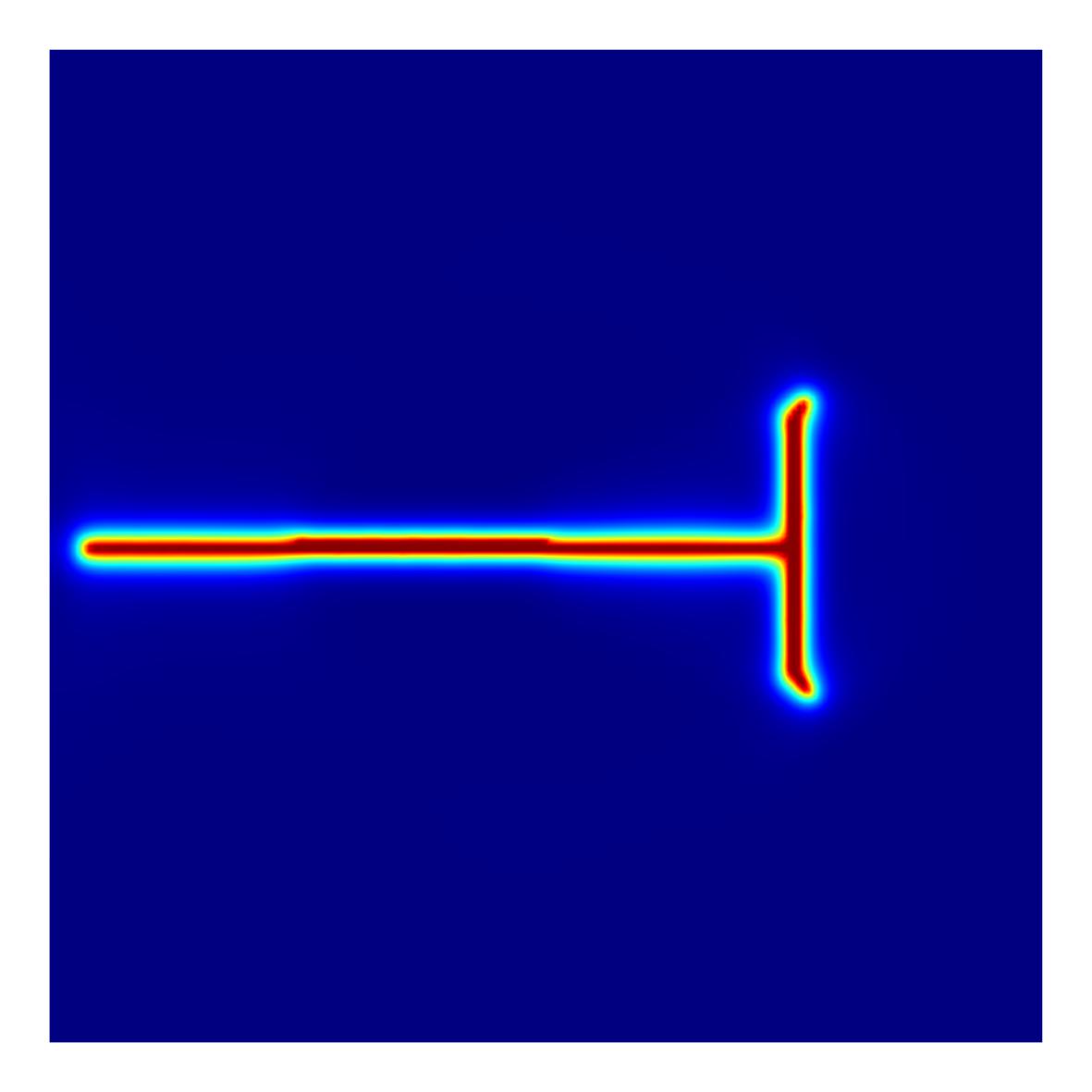}}
	\subfigure[$t=11.6$ s]{\includegraphics[height = 5cm]{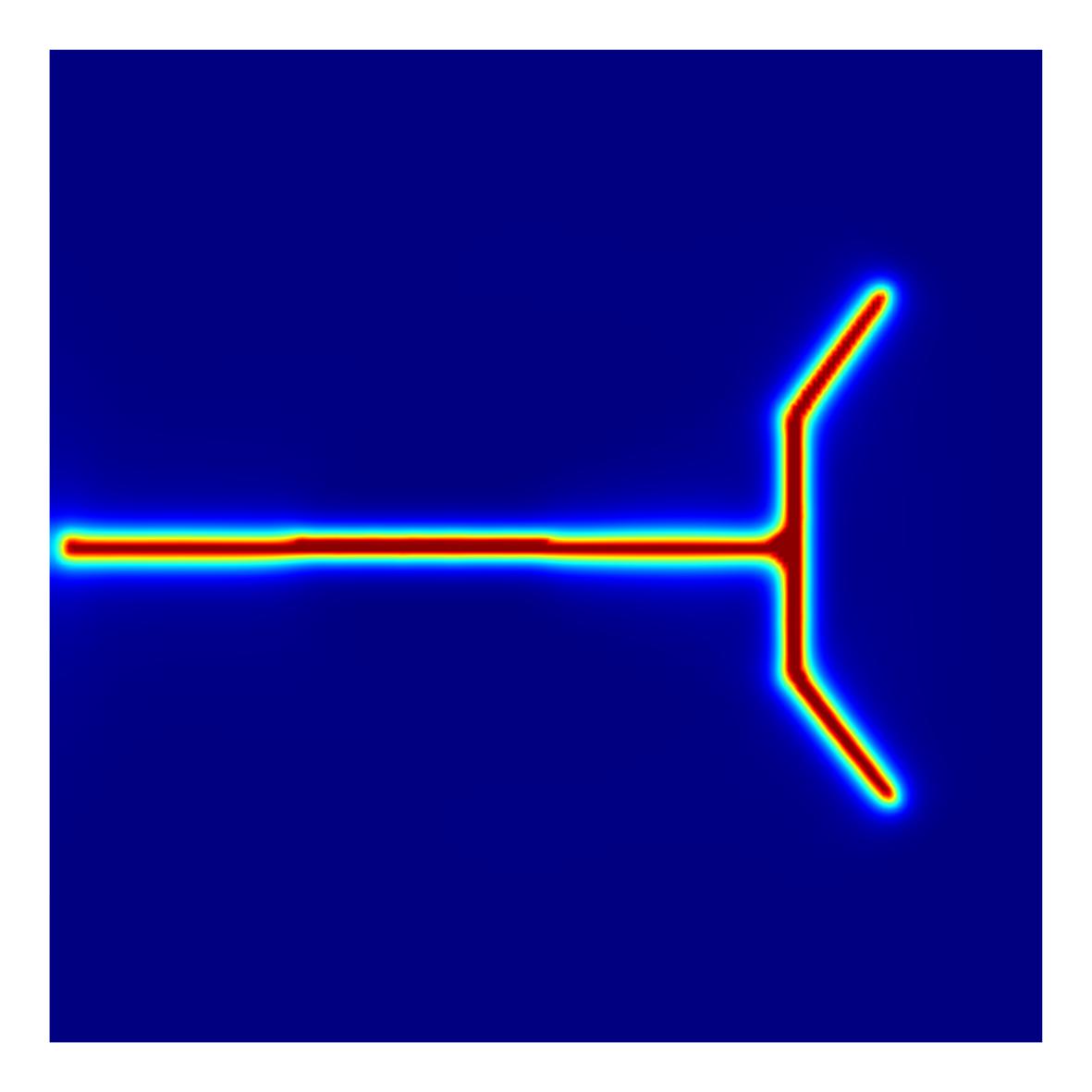}}
	\caption{Crack patterns for the example of two perpendicular pre-existing cracks}
	\label{Crack patterns for the example of two perpendicular pre-existing cracks}
	\end{figure}

\begin{figure}[htbp]
	\centering
	\subfigure[$t=0$]{\includegraphics[height = 5cm]{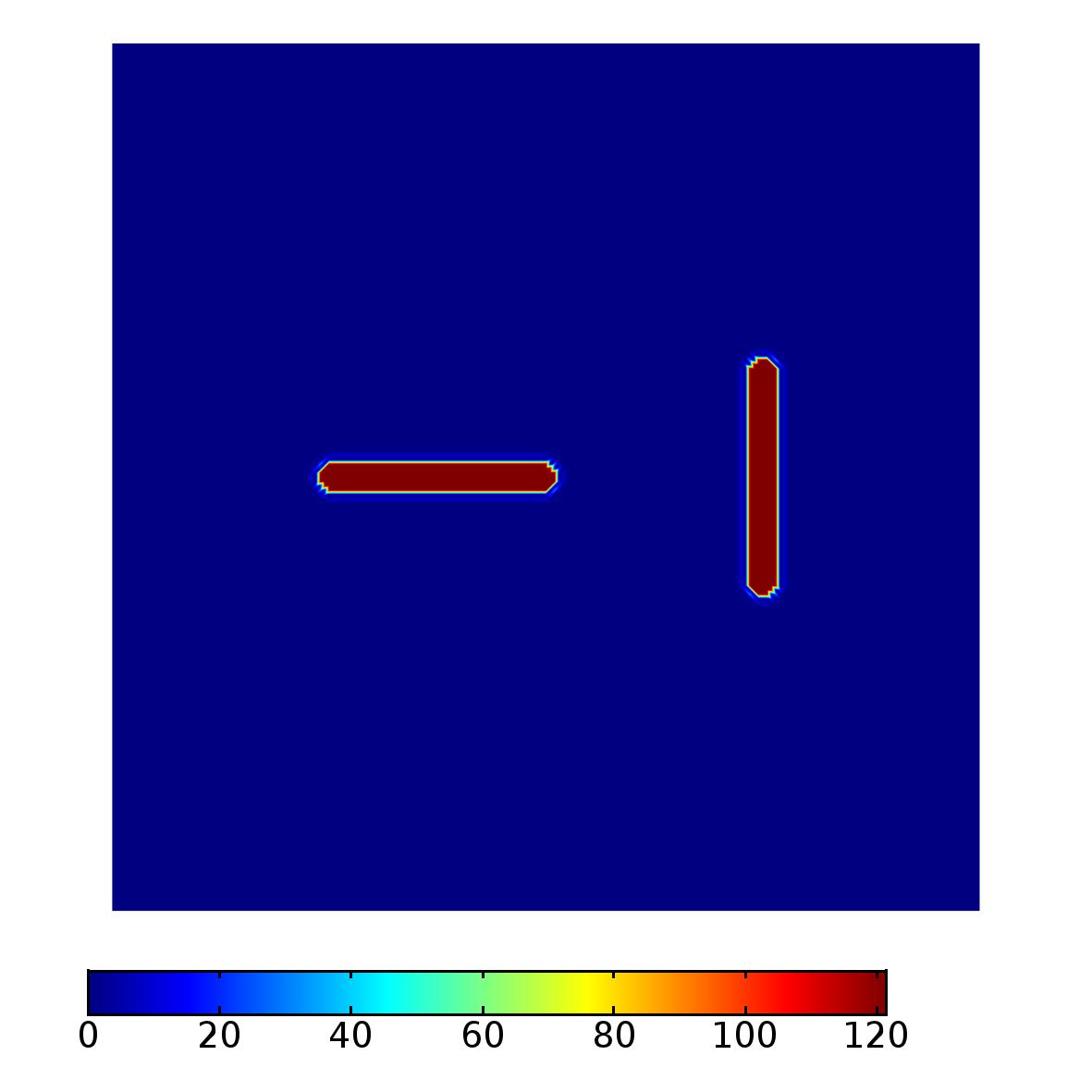}}
	\subfigure[$t=6$ s]{\includegraphics[height = 5cm]{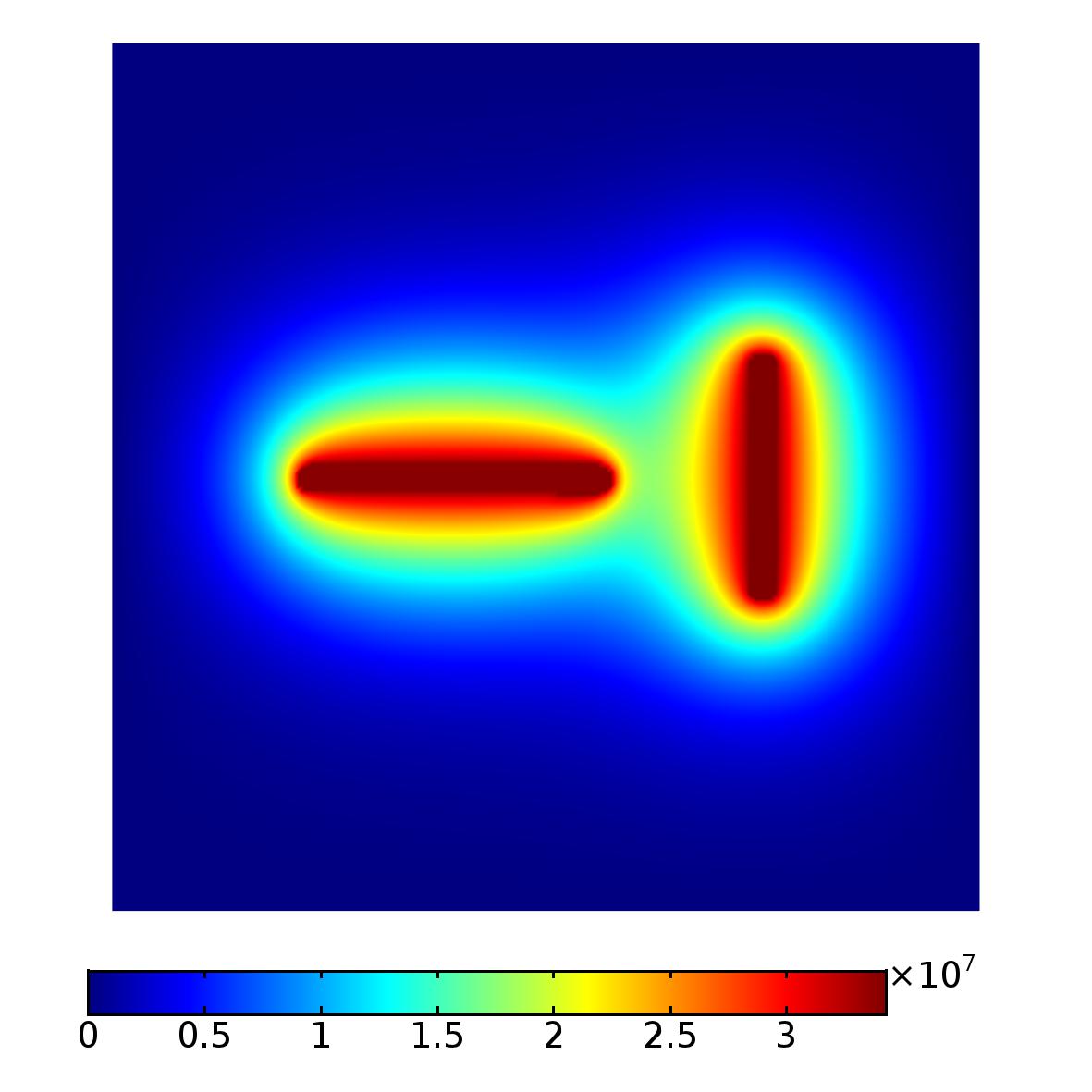}}
	\subfigure[$t=6.8$ s]{\includegraphics[height = 5cm]{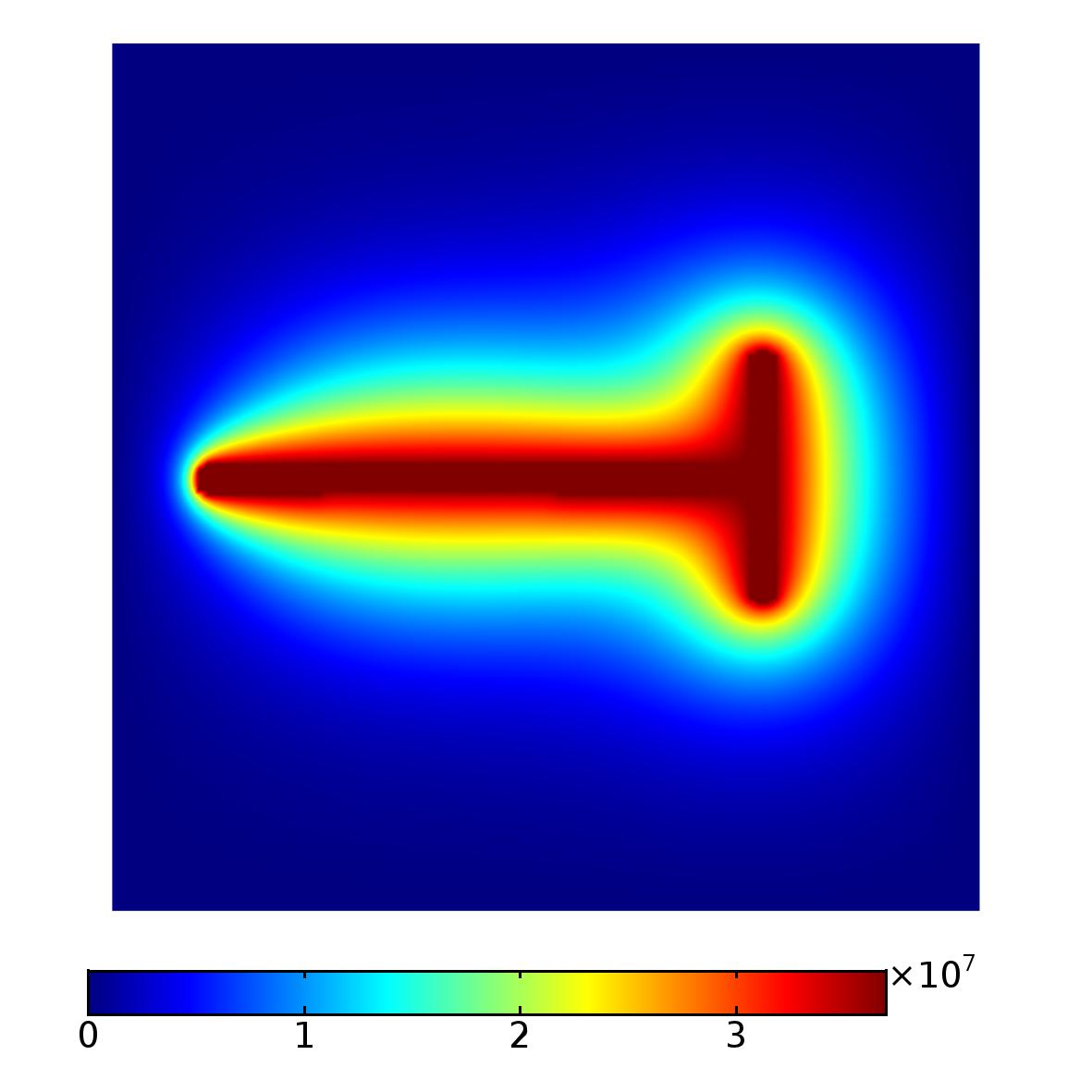}}\\
	\subfigure[$t=8$ s]{\includegraphics[height = 5cm]{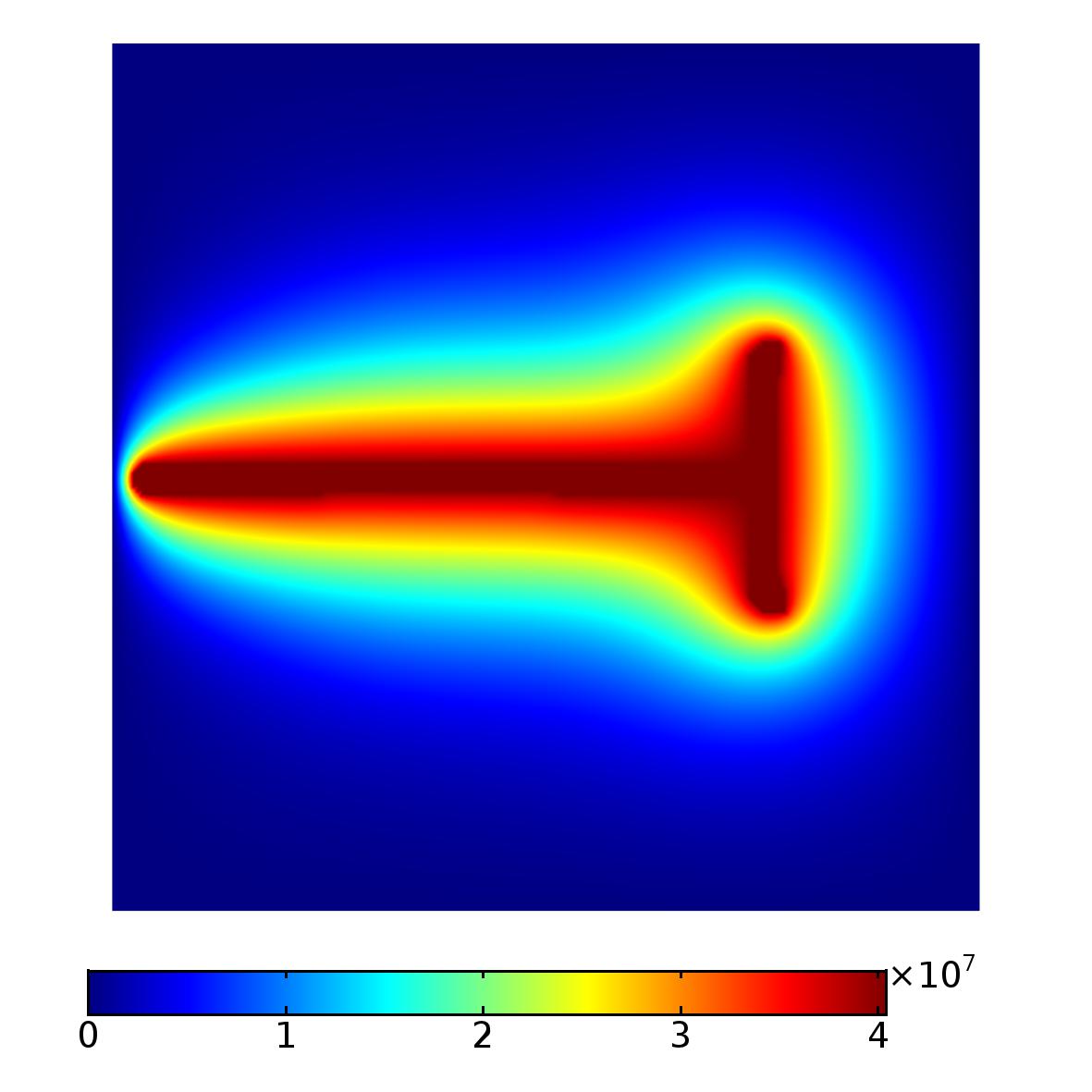}}
	\subfigure[$t=11.6$ s]{\includegraphics[height = 5cm]{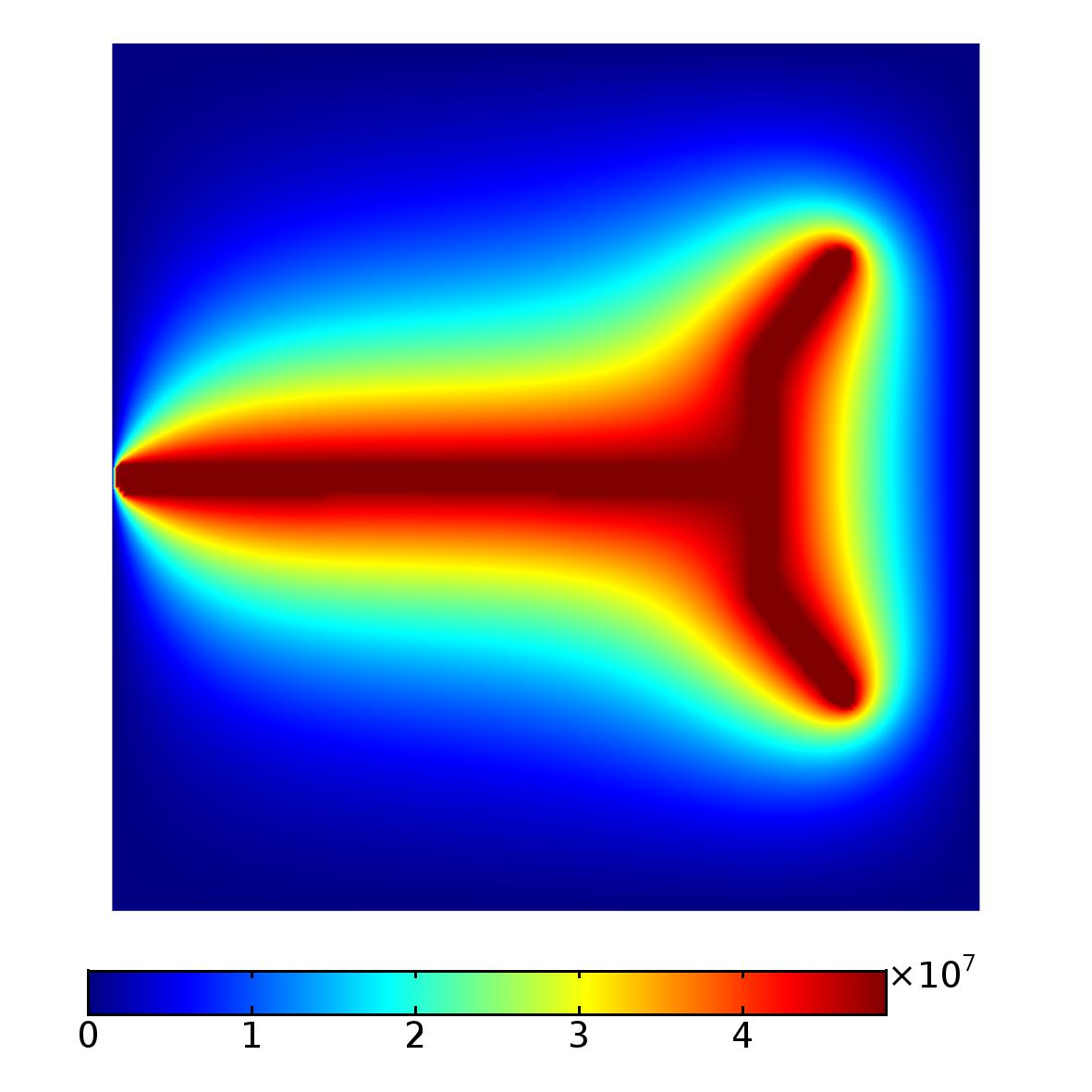}}
	\caption{Pressure distribution for the example of two perpendicular pre-existing cracks}
	\label{Pressure distribution for the example of two perpendicular pre-existing cracks}
	\end{figure}

\begin{figure}[htbp]
	\centering
	\includegraphics[width = 8cm]{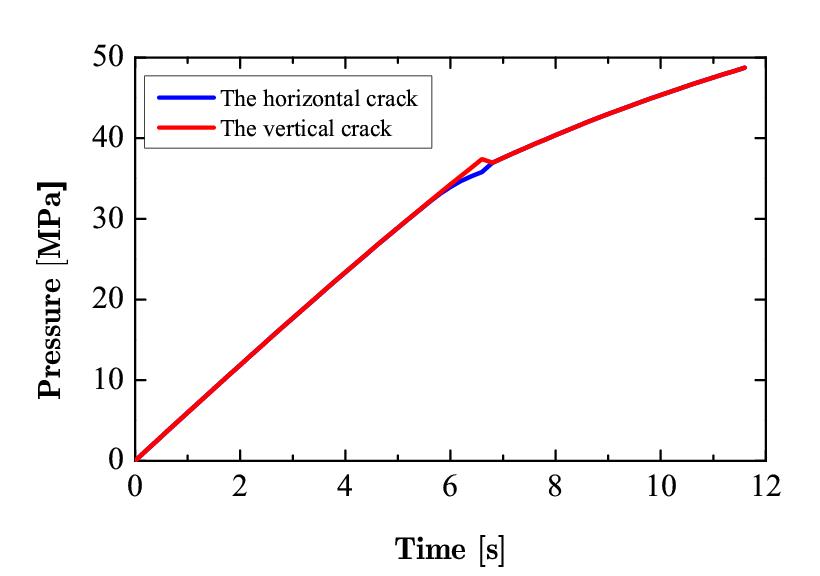}
	\caption{Pressure-time curve  for the example of two  perpendicular pre-existing cracks}
	\label{Pressure-time curve  for the example of two perpendicular pre-existing cracks}
	\end{figure}

Figure \ref{Crack patterns for the example of two perpendicular pre-existing cracks} shows the evolution of the phase field. As the time increases, only the horizontal crack propagates in the early period while there is no propagation for the vertical crack. When $t=$ 6.8 s, the horizontal and vertical cracks coalesce. Subsquently, the horizontal crack propagates to the left boundary of the specimen and the vertical crack propagates at a large angle to the horizontal direction. Figure \ref{Pressure distribution for the example of two perpendicular pre-existing cracks} presents the pressure distribution for different times and Fig. \ref{Pressure-time curve  for the example of two perpendicular pre-existing cracks} depicts the pressure-time curve  for the two perpendicular pre-existing cracks. The pressure of the two cracks is almost the same and only little fluctuations occur when the cracks coalesce.

\subsection{Propagtion of two parallel penny-shaped cracks in 3D}

In the last example, we demonstrate the capability of the proposed approach for 3D problems. We set two parallel penny-shaped cracks in the domain of (-2 m, 2 m)$^3$. The two initial cracks are centered at (0, 0, 0.3 m) and (0, 0, -0.3 m) and on the planes of $y=0.3$ m and $y=-0.3$ m, respectively. Both cracks have the same initial radius of 0.5 m. We set the length scale parameter to $l_0 = 0.06$ m. The maximum element size $h$ is set as 0.06 m and a total of 774,480 6-node prism elements are used to discretize the domain. The time increment is set as $\Delta t=0.01$ s and 1330 time steps are totally used with the time cost of 175 h 30 min on four CPUs.

Crack propagation patterns of the two parallel penny-shape cracks in 3D are shown in Fig. \ref{Crack patterns for the example of  two parallel penny-shaped cracks in 3D}. The radii of the cracks become more than 0.5 m when $t=13.2$ s. However, when $t=13.3$ s, the cracks propagate to bowl-shaped volumes. The 3D crack patterns obtained by the proposed approach are in good agreement with those reported in \citet{lee2016pressure}. However, only a coarse mesh used in the 3D simulation and no extra re-meshing and adaptive techniques are adopted in this paper. This fully indicates the strong capability of phase field models to simulate hydraulic fracturing.

	\begin{figure}[htbp]
	\centering
	\subfigure[$t=0$]{\includegraphics[height = 6cm]{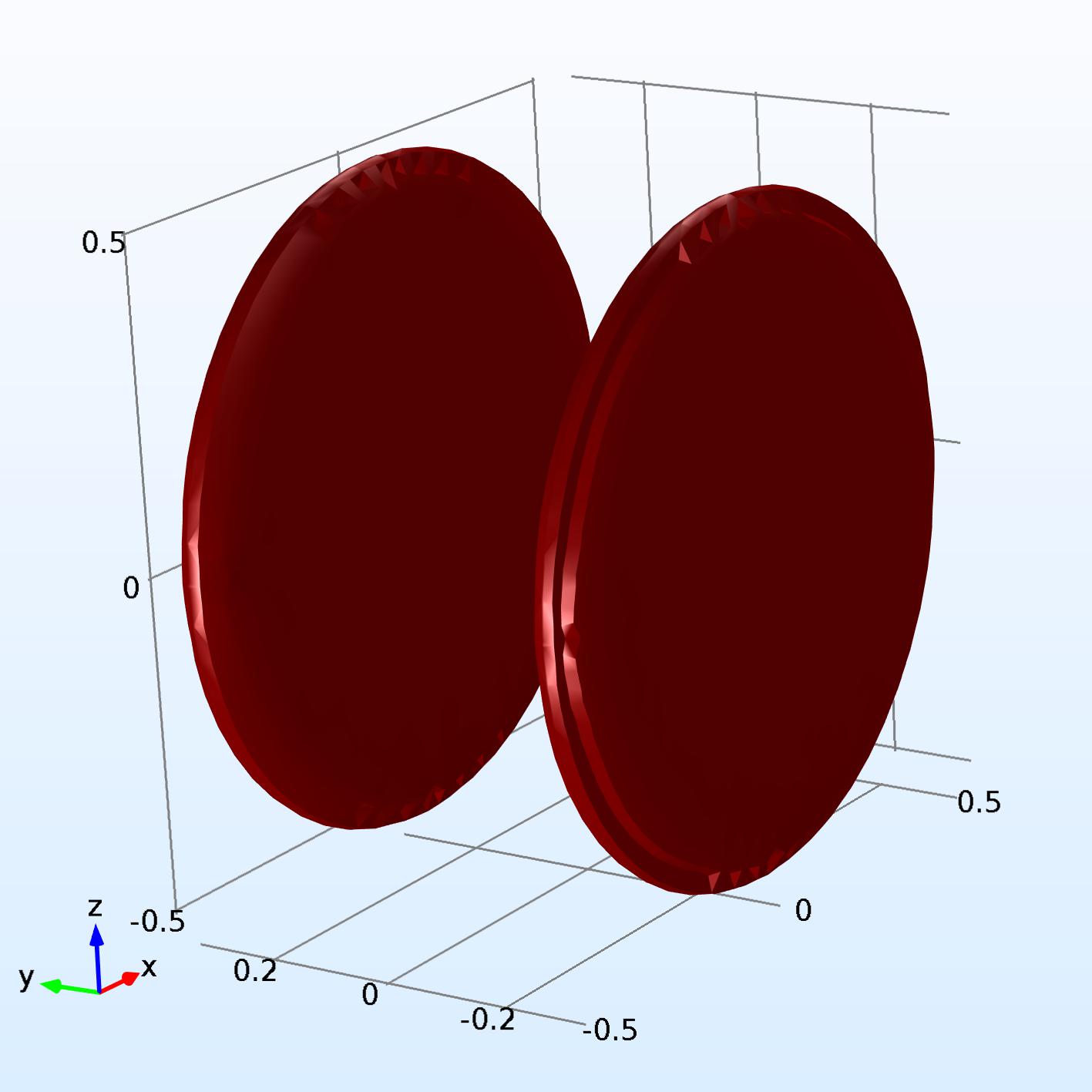}}
	\subfigure[$t=13.2$ s]{\includegraphics[height = 6cm]{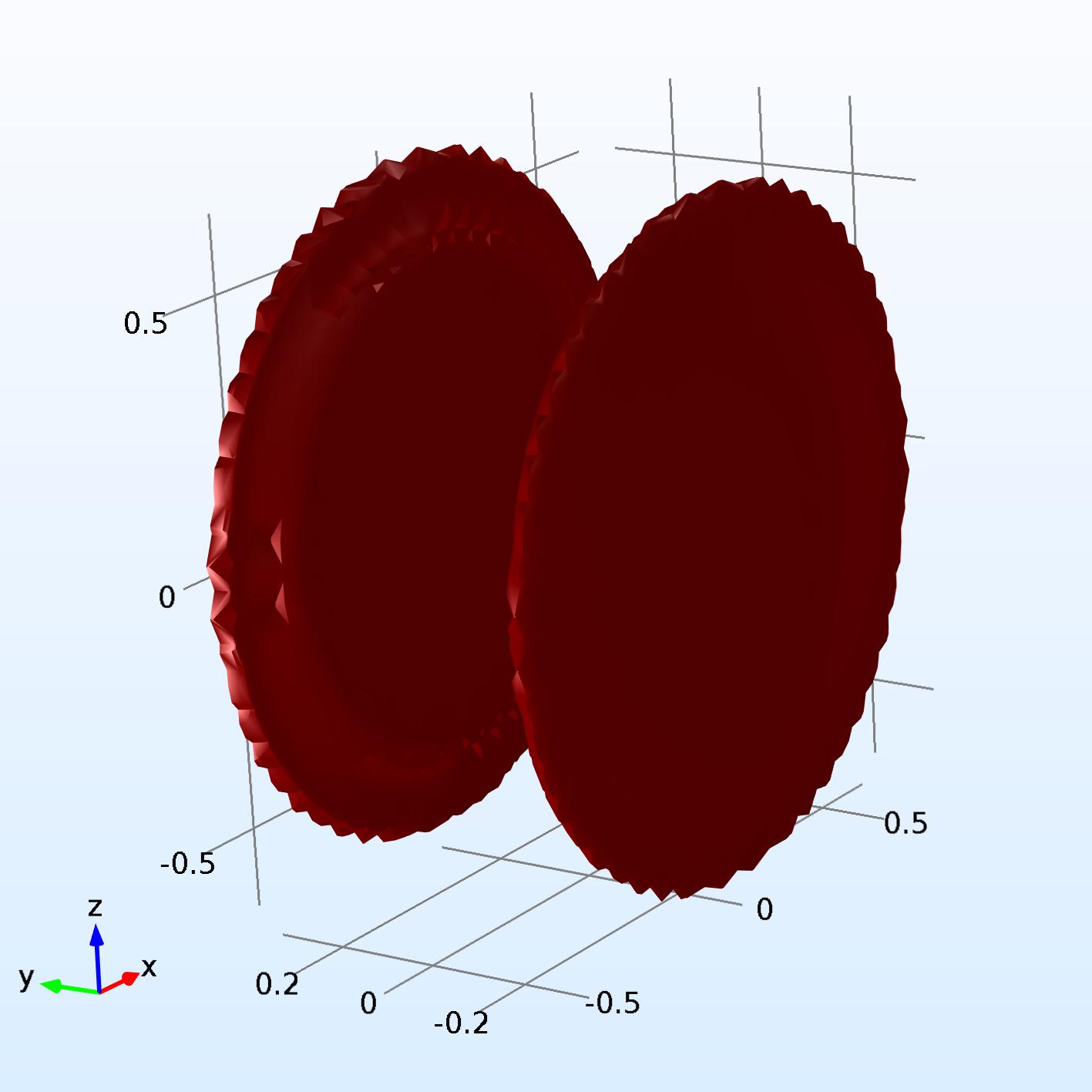}}\\
	\subfigure[$t=13.3$ s]{\includegraphics[height = 6cm]{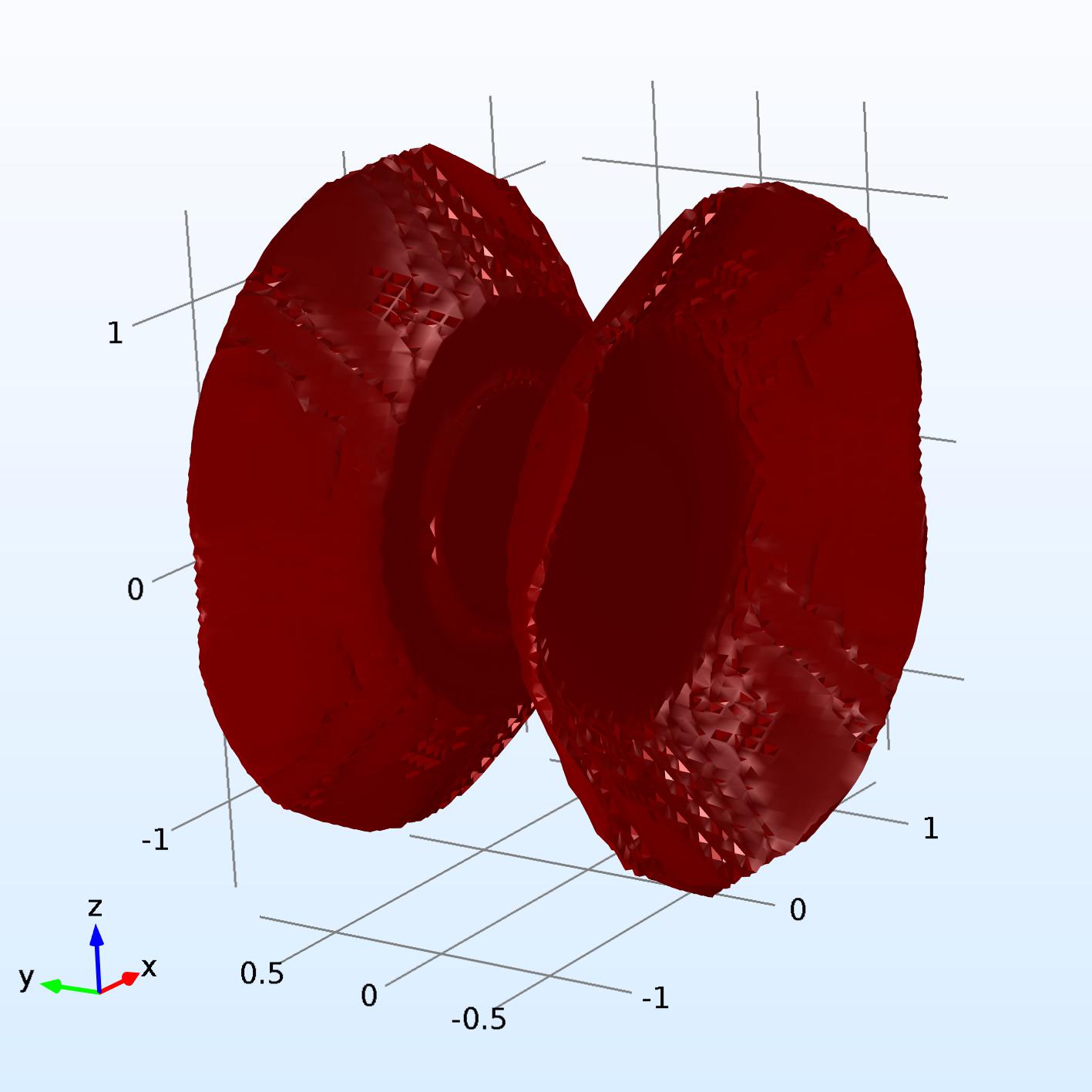}}
	\caption{Crack patterns for the example of two parallel penny-shaped cracks in 3D}
	\label{Crack patterns for the example of two parallel  penny-shaped cracks in 3D}
	\end{figure}

\section {Conclusions}

A phase field model for fracture in poroelastic media is proposed in this paper. The porous medium is modeled based on the classical Biot poroelasticity theory and the fracture behavior is captured by the phase field model. The fracture propagation is driven by the elastic energy and the phase field is used as an interpolation function to transit fluid property from the intact medium to the fully broken one. The model is implemented in Comsol Multiphysics and solved by a segregated scheme. The proposed approach is verified for two benchmark problems: a single-phasic solid subjected to tension and a 2D specimen subjected to an increasing internal pressure. Results are compared to analytical solutions. Finally, 2D and 3D examples of  internal fluid injection are presented to show capability of the proposed approach in hydraulic fracturing. 

The proposed phase field approach overcomes the disadvantages of previous phase field approaches in HF. Taking the computational cost as an example, \citet{yoshioka2016variational} used up to 100 CPUs to run a general 3D example within several days while we run the 3D example in this paper in a week only by 4 CPUs. Moreover, we give full implementation details of the approach and make the implementation fully available in Comsol Multiphysics. The results presented in this work can be also extended to inelastic, partially saturated, or heterogeneous porous media and multiple complex fractures.  

\section*{Acknowledgement}
The financial support provided by the Sino-German (CSC-DAAD) Postdoc Scholarship Program 2016, the Natural Science Foundation of China (51474157), and RISE-project BESTOFRAC (734370) is gratefully acknowledged.

\bibliography{references}

\end{document}